\documentclass[pre,showpacs,amsmath,amssymb,twocolumn]{revtex4}

\usepackage{graphicx}
\usepackage{bm}

\begin{document}

\title{Classical Nonlinear Response of a Chaotic System: Collective Resonances}

\author{Sergey V.  \surname{Malinin}$^a$}
\author {Vladimir Y. \surname{Chernyak}$^a$}
\email{chernyak@chem.wayne.edu}
\affiliation{$^a$Department of Chemistry, Wayne
State University, 5101 Cass Ave, Detroit, MI 48202\\}
\date{\today}

\begin{abstract}
We consider the classical response in
a chaotic system. In contrast to behavior in integrable or almost integrable systems,
the nonlinear classical response in a chaotic system vanishes at long
times. The response also reveals certain features of collective resonances which do not correspond
to any periodic classical trajectories. The convergence of the response is shown
to hold due to the exponential
time dependence of the stability matrix. The growing exponentials corresponding to strong
instability do not inhibit the convergence.
We calculate both linear and second-order response in one of the simplest chaotic
systems: free classical motion on a surface of constant negative curvature.
We demonstrate the relevance of the model for applications to spectroscopic experiments.

\end{abstract}

\maketitle

\section{Introduction}

Time-resolved femtosecond spectroscopy constitutes a powerful tool that probes electronic and
vibrational coherent dynamics of complex molecular systems in condensed phase
\cite{TokmakoffetalPRL,ZanniHochstrPNAS,Fayer01,Jonas03,StolowJonas,Ashbury}.
Spectroscopic techniques allow for direct measurements of the time- and frequency-domain
optical response functions that carry detailed information on the underlying
dynamical phenomena. Nonlinear spectroscopy contains additional information on nonequilibrium
processes absent in linear measurements.

In complex systems, such as polyatomic molecules, vibrational dynamics of many degrees
of freedom can be adequately
described in the framework of classical mechanics.
In particular, at energies corresponding to room temperatures the complexity often originates
from heavy strongly anharmonic vibrational modes that can be treated classically.

A number of studies have been devoted to the nonlinear response in stable (integrable) dynamical
systems \cite{LeegwaterMukamel95,NEL04,NL05,KryvohuzCaoPRL05,KryvohuzCaoPRL06}. Classical nonlinear
response functions have been shown to diverge with time in the general case
\cite{KryvohuzCaoPRL06}, although in certain cases the divergence can be removed by thermal
averaging using the canonical distribution \cite{LeegwaterMukamel95,NEL04}.
The latter effect can be viewed as a result of destructive interference of different paths
(dephasing), while the divergence is associated with the rephasing.

The divergence would pose a major problem for the whole concept of perturbative response. On the
other hand, quantum response functions do not exhibit any divergences. However, application of the
fully quantum description is often unmanageable in the systems of interest. It has been
demonstrated that systematic $\hbar$ expansion cures the divergence when the limit of vanishing
$\hbar$ is taken after obtaining the asymptotics of long times \cite{KryvohuzCaoPRL05,NL05}.
Alternatively the divergence can be removed by adding noise that models interaction with a bath (or
a solvent) \cite{GojLoring}.

Previous analytical studies of the nonlinear response were mostly limited to fully solvable models,
which implies integrable dynamics in the classical limit. Integrable dynamics corresponds to an
idealized picture for a realistic physical situation, since even weak perturbations generally
destroy integrability. However, the problem is that weak deviations from the integrability do not
eliminate the unphysical behavior of the response functions. This follows from the fact that the
divergence in integrable systems originates from quasiperiodic motions on invariant tori
\cite{KryvohuzCaoPRL05}. According to the Kolmogorov-Arnold-Moser (KAM) theory, most invariant tori
are not destroyed by small perturbations that break down integrability \cite{Arnold}. Therefore,
the long-time behavior of the response in almost integrable systems is similar to its integrable
counterpart.

Stable dynamics is typical for a situation close to equilibrium. At larger energies a generic
situation corresponds to dynamical behavior with chaotic features \cite{Gutzwiller}. Moreover,
unstable (hyperbolic) dynamics may be more common due to the stability of chaos with respect to
perturbations. It has been argued based on results of numerical analysis \cite{DellagoMukamel03}
that chaotic dynamics appears to observe the convergence of the classical response functions. In
spite of apparent importance and to the best of our knowledge, the problem of the nonlinear
response in strongly chaotic systems has never been addressed using analytical methods. We note
that analytical calculations of the response are rarely feasible in nonintegrable systems, whereas
numerical simulations of chaotic dynamics are complicated by the exponential divergence of
stability matrices \cite{DellagoMukamel03}.

A strongly chaotic (mixing) system is characterized by a special spectrum of the Liouville operator
\cite{Pollicott,Ruelle86,RobertsMuz}. The spectrum consists of complex Ruelle-Pollicott (RP)
resonances that
determine the asymptotic oscillations and decay of the correlations. This yields the linear
response function directly related to the two-point correlation functions, in agreement with the
fluctuation dissipation theorem (FDT). Nonlinear response, however, turns out to be more involved
still with the noticeable effect of the resonances.

In this work we show that (i) the classical linear and nonlinear response of a chaotic system
exhibits decay and oscillations as a function of times between the driving pulses, and (ii) the
Fourier transform of $2D$ second-order response function reveals broad and asymmetric peaks that
can be viewed as signatures of chaos in underlying dynamics.

We first establish a general qualitative picture of linear and nonlinear response in a classical
system with hyperbolic chaotic dynamics and demonstrate that the classical response functions
exponentially vanish at large times. To exemplify our arguments we further perform detailed
analytical calculations of linear and second-order response functions for a free particle moving
along a compact surface of constant negative curvature. The model, that constitutes a well-known
example of classical chaos and  a prototype for quantum chaos, allows for an exact solution due to
strong dynamical symmetry. To the best of our knowledge, we present the first explicit analytical
calculation of nonlinear response in a chaotic system.

The manuscript is organized as follows. We begin with reviewing the classical response theory using
the Liouville representation of classical mechanics.
In Section \ref{section:qualitative} we present the qualitative picture for the classical nonlinear
response functions and argue that they show exponential decay
at long times. We also argue that a generic finite motion of the molecular system with potential interactions
is similar to the free motion along an effective compact configuration space of negative
curvature. In Section
\ref{section:free-partcile-2D} we consider the chaotic model
of free motion along a Riemann surface of constant negative curvature. We demonstrate how to make
use of strong dynamical symmetry, and reduce the model to a much simpler problem of $1D$ dynamics
on a circle. In Section \ref{section:calculation} we present our calculation of the linear and
second-order response functions. This calculation constitutes the main technical result of the
paper. The results of numerical calculations of $2D$ spectroscopic signals are presented in Section
\ref{section:numerical}. All necessary technical details are summarized in the Appendices.

\section{Classical response in Liouville space}
\label{section:classical-liouville}

In this section we review a general formalism of the classical response theory, using the language
of probability distributions (e.g. see Ref. \onlinecite{DellagoMukamel03}).
We adopt the Liouville picture of
classical mechanics where the system state that evolves in time is described by a distribution
$\rho({\bm \eta},t)$, whereas the observables are represented by functions in phase space. This
language is advantageous in a chaotic case, since mixing (e.g., hyperbolic) dynamics is
characterized by strong instability in trajectory space, and individual trajectories do not provide
meaningful information on the system relaxation.

Consider classical Hamilton dynamics that occurs in phase space $M$ equipped with a Poisson
bracket. Evolution of an externally driven system is described by the classical Liouville equation
with a time-dependent Hamiltonian $H_{T}(t)$
\begin{eqnarray}
\label{Liouville-equation} \frac{\partial\rho({\bm \eta},t)}{\partial t}=-\hat{L}_{T}(t)\rho({\bm
\eta},t)\,,
\\
\hat L_{T}(t)\ldots=\left\{H_{T}(t),\ldots\right\}\,, \quad H_{T}(t)=H-{\cal E}(t)f\,,
\end{eqnarray}
where $H({\bm \eta})$ is the time independent Hamiltonian of the undriven system, ${\cal E}(t)$ is
the time dependent uniform driving field, and dipole moment (polarization) $f({\bm \eta})$
describes the coupling of the system to the driving field. This sign choice of in the Poisson
bracket $\{f,g\}$ corresponds to a convention that
the action of the Liouville operator $\hat{L}_{T}$ determines the phase space velocity as $\dot{\bm
\eta} = \hat L_T(t){\bm \eta}$.

In many cases $f({\bm \eta})$ is also an observable, e.g. polarization that creates spectroscopic
signals, and the measured value is given by the time dependent average
\begin{eqnarray}
\langle f\rangle=\int d\bm\eta\,\rho({\bm \eta},t) f({\bm \eta})\,.
\end{eqnarray}
The undriven system with Hamiltonian $H$ is characterized by equilibrium distribution
$\rho_0(\bm\eta)$ that depends only on the conserved energy $H$ of the system, i.e.
$\{H,\rho_0\}=0$. The average of $f$ for the equilibrium distribution $\rho_0(\bm\eta)$  normally
vanishes (in spectroscopic literature this is often referred to as the absence of the permanent
dipole):
\begin{eqnarray}
\int d\bm \eta\,\rho_0(\bm\eta) f(\bm\eta)=0\,.
\end{eqnarray}

The response theory is usually based on a perturbative expansion of the measured signal in powers
of the driving field. The response functions can be naturally obtained via calculating the
correction to the equilibrium density distribution due to the driving field. The full Liouville
equation is repeatedly solved, this results in an expansion $\rho=\rho_0+\rho_1+\ldots$ where
$\rho_n$ represents the contribution of $n$-th order in the field ${\cal E}$.
Therefore, we start with solving the equation
\begin{eqnarray}
\label{Liouville-equation-correction} (\partial_t+\hat L)\rho_1={\cal E}(t)\{f,\rho_0\}\,,
\end{eqnarray}
to get the linear in ${\cal E}$ correction to the density distribution:
\begin{eqnarray}
\rho_1(t)=\int\limits_0^t d\tau_1 \,{\cal E}(\tau_1)e^{-\hat L(t-\tau_1)}\{f,\rho_0\}\,.
\end{eqnarray}
Here $\hat L$ is the Liouville operator
\begin{eqnarray}
\label{define-Liouville} \hat L \rho=\left\{H,\rho\right\}
\end{eqnarray}
of the unperturbed dynamics that corresponds to the Hamiltonian $H$. The corrections to $\rho_0$ in
all orders of ${\cal E}$ can be obtained by solving equations similar to Eq.
(\ref{Liouville-equation-correction}).
This leads to the natural representation of the observable quantity in the form of an expansion
\begin{eqnarray}
\langle f\rangle \equiv \int d{\bm\eta}\, \rho({\bm\eta},t) f({\bm\eta}) = \int\limits_0^t d\tau_1
\,S_1(t;\tau_1){\cal E}(\tau_1)+
\\
\int\limits_0^t d\tau_2 \int\limits_0^{\tau_2} d\tau_1 \,S_2(t;\tau_1,\tau_2) {\cal E}(\tau_1){\cal
E}(\tau_2)+\dots\,,
\end{eqnarray}
which defines response functions $S_n(t;\dots,\tau_1)$. The response function of order $n$ depends
on $n+1$ time variables, it can be reduced to $n$ time segments $t_n=t-\tau_n, \dots,
t_1=\tau_2-\tau_1$ in the case when the unperturbed dynamics has no explicit time dependence. For
time segments representation, the linear and second-order response functions read:
\begin{align}
\label{first-order} S^{(1)}(t_1) & =\int d{\bm \eta}\, f({\bm \eta})e^{-\hat L t_1}\{f({\bm
\eta}),\rho_0\}\,,
\\
\label{second-order} S^{(2)}(t_1,t_2) & =\int d{\bm \eta}\, f({\bm \eta})e^{-\hat L t_2} \{f({\bm
\eta}),e^{-\hat L t_1}\{f({\bm \eta}),\rho_0\}\}\,.
\end{align}
These expressions for the first- and second-order response will be the starting point of our
general qualitative analysis of response functions in hyperbolic systems. They will be also
utilized in our analytical calculations for a particular chaotic system where strong dynamical
symmetry of the phase space enables us to solve the problem completely.

\section{Qualitative Picture of Response in a Hyperbolic Dynamical System}
\label{section:qualitative}

\subsection{Integrable and almost integrable dynamics}
\label{subsection:integrable}

Dynamics of a classical Hamiltonian system in the vicinity of an equilibrium (stationary) point can
be adequately described by a system of uncoupled harmonic oscillators. This is achieved by
expanding the classical Hamiltonian $H$ up to second order in dynamical variables (the first-order
terms vanish for an expansion around a stationary point). A harmonic system represents the
simplest example of integrable dynamics.

The evolution of integrable systems with $n$ degrees of freedom is naturally described in terms of
$n$ canonical pairs of action and angle variables. Action variables $c_{j}$ with $j=1,\ldots,n$ are
first integrals of motion $\left\{H,c_{j}\right\}=0$ in involution, i.e.
$\left\{c_{i},c_{j}\right\}=0$, with the Hamiltonian $H({\bm\eta})=H(c_{1},\ldots,c_{n})$ depending
on the integrals of motion only. The corresponding phase space vector fields $\hat{c}_{j}$ defined
by $\hat{c}_{j}f=\{c_{j},f\}$, therefore, commute and describe motions along the tori with
frequencies $\omega_{j}=\partial H(c_{1},\ldots,c_{n})/\partial c_{j}$
that depend parametrically on the integrals of motion.
Since for fixed values of the integrals of motion the angular velocities are constant, such
dynamics is referred to as is conditionally periodic (or quasi-periodic) motion. For given values
of action variables determined by initial conditions, the trajectories lie on the corresponding
$n$-dimensional torus which is a subspace of $2n$-dimensional phase space. The motion on the torus
is strictly periodic if the ratios of corresponding angular velocities are rational (resonant
torus); if the angular velocities are incommensurate, the trajectories densely cover the torus
(non-resonant torus). Harmonic dynamics represents the simplest particular case of integrable
dynamics when the frequencies $\omega_{i}$, $i=1,\ldots,n$ are independent of the action variables
$c_{j}$.

The nonlinear response function for classical integrable dynamics have been shown
to have power-like divergence
at long times
which can be eliminated by invoking a quantum description
\cite{LeegwaterMukamel95,NEL04,NL05,KryvohuzCaoPRL05,KryvohuzCaoPRL06}.
Divergence of nonlinear response in integrable
systems does not imply unphysical behavior by itself, since integrable dynamics is an idealization.
In real systems quantum effects, interactions with the bath or irregular dynamics may provide
a necessary regularization. While the first two phenomena have been discussed in the literature,
we concentrate on the latter.
A small, yet generic, perturbation of the system Hamiltonian destroys integrability.
However such a perturbation does not break down
the linear divergence in the response functions.
This follows from the qualitative picture of
almost integrable dynamics established by the celebrated Kolmogorov-Arnold-Moser (KAM) perturbation
theory \cite{Arnold}. The  KAM theory states that a sufficiently small perturbation does not
destroy most nonresonant tori,
which means that in the invariant
subspace of the entire phase space represented by the remaining distorted tori the
dynamics preserves its quasiperiodic nature.
Although motion inside the instability zones represented by
destroyed tori becomes chaotic, their relative measure in phase space is small for small
perturbations to the integrable dynamics. In the simplest $n=2$ case the zones of instability are
confined between remaining invariant tori.

It has been pointed out \cite{NEL04,KryvohuzCaoPRL05} that the divergence or the classical response
functions in integrable systems originates from quasiperiodic nature of the underlying motion.
Combined with the picture of almost integrable dynamics established by KAM theory, this demonstrates
the unphysical divergence of the response functions is stable with respect to at least weak
deviations from integrability.
We will see that chaotic regions (instability zones) do not
contribute to the divergence,
therefore the diverging terms will decrease with increasing the deviation from
the integrable situation due to decrease of the amount of surviving invariant tori.

\subsection{Chaotic dynamics}
\label{subsection:hyperbolic}

Integrable and almost integrable Hamiltonian dynamics considered in subsection
\ref{subsection:integrable} is typical for low energies when the system is moving in the
neighborhood of a stable stationary  point (equilibrium).
Islands of stable dynamics completely vanish for larger deviations from the integrability
caused by higher characteristic energies.
The motion will be typically chaotic for higher temperatures or in essentially nonequilibrium
processes such as photoinduced molecular dynamics on the excited electronic adiabatic surfaces.

In this subsection we present a qualitative picture of nonlinear response in chaotic systems.
Due to FDT, the
stability matrices were found to affect the response functions starting with the second
order \cite{MKC96,DellagoMukamel03}.
Therefore, the
divergence of the nonlinear response functions in the integrable case
can be attributed to the growth with time of certain
stability matrix components. In the case of strong chaos there are components of the stability
matrices that grow exponentially $\sim e^{\lambda t}$ with time, $\lambda$ being the Lyapunov exponent.
Numerical simulations have demonstrated that at least for some examples of chaotic dynamical
systems the classical nonlinear response functions are free of unphysical divergences
\cite{DellagoMukamel03}.

In this subsection we present qualitative arguments that rationalize physical
exponentially decaying at long times
behavior of classical nonlinear response functions in systems
that exhibit strong enough chaotic behavior. More precisely we will consider mixing systems also
known as A-flows, Smale or uniformly hyperbolic dynamical systems (for a nice overview without too
many details see Ref. \onlinecite{Ruelle99}). In these systems at all points of the $(2n-1)$-dimensional
energy shell, or all relevant points (the relevant points belong to the so-called
nonwandering subset) in the Smale's case, one can define $n_{+}$ unstable and $n_{-}$ stable
tangent directions with $n_{+}+n_{-}=2n-2$, so that a small deviation from a trajectory
along the stable (unstable)
directions decays exponentially in time for forward (backward) dynamics. The stable (unstable)
directions can be locally integrated to obtain stable (unstable) manifolds of dimension $n_{-}$
($n_{+}$). Note that $n_{+}$ ($n_{-}$) is the number of positive (negative) Lyapunov exponents. Also
note that in the Hamiltonian dynamics case the volume in phase space is conserved
so that all Lyapunov exponents sum up to zero
\begin{eqnarray}
\label{Lyapunov-sum}
\lambda\equiv\sum_{j=1}^{n_{+}}\lambda_{j}^{(+)}=-\sum_{j=1}^{n_{-}}\lambda_{j}^{(-)}\,.
\end{eqnarray}

We will consider the simplest case of the lowest dimension coordinate space dimension $n=2$ that
allows for chaotic dynamics. The isoenergetic shell is $3$-dimensional, we have $n_{+}=n_{-}=1$,
and two nontrivial Lyapunov exponents $\pm\lambda$.

\begin{figure}[ht]
\centerline{
\includegraphics[width=2.8in]{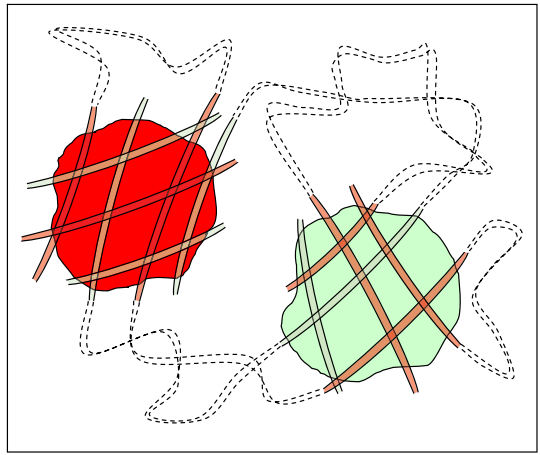}
}
\caption{
Schematic picture of
the cross-section of the phase space along the surface defined by the stable and
unstable directions.
Initial distribution of $f$ is presented by
two regions oppositely charged regions (dark red and light green).
As time elapses the distributions elongate along unstable directions
and contract along stable ones.
\label{fettuccine}
}
\end{figure}

A schematic picture of first- and second-order response formation in a hyperbolic chaotic system is
shown in Fig.~\ref{fettuccine}. Since chaotic dynamics described in terms of phase-space
trajectories is extremely unstable, using the dual representation via evolution of distributions
(Liouville representation) will be advantageous. We start with the linear response function
$S^{(1)}(t)$.
Since the equilibrium distribution depends on the system energy only, we can recast the expression
for the linear response [Eq.~(\ref{first-order})] in a form that actually represents the FDT
\begin{eqnarray}
\label{S-1-FDT} S^{(1)}(t)=\partial_{t}I^{(1)}(t)=\partial_{t}\int d{\bm\eta}f
e^{-\hat{L}t}f\partial_{E}\rho_{0}\,.
\end{eqnarray}
Since $\int d{\bm x} f=0$ (no permanent dipole), we can consider the simplest set-up when the
function $f$ is represented by two small separate regions in the phase space where it adopts
positive and negative values (an extension of our arguments to the general case is
straightforward). By our assumption the linear sizes of the regions $a\ll l$ are small compared to
the linear size $l$ of the compact phase space.
Evolution during time $t$ changes the shapes of the regions.
In the case of hyperbolic dynamics for $t\gg 1$, the shape becomes similar to ribbon-like fettuccine:
elongated along the unstable direction by a factor $e^{\lambda t}$, narrowed along the stable one
by $e^{-\lambda t}$ (we reiterate that $\lambda$ is the positive the Lyapunov exponent) and
unchanged along the flow.
The long ribbon becomes evenly folded in the entire accessible energy shell.
The overlap of the distributions $f$ and
$e^{-\hat{L}t}f\partial_{E}\rho_{0}$ in Eq.~(\ref{S-1-FDT}) is represented by a large number
$N_{1}(t)$ of disconnected regions.
The dimensions of each region are typically $a$, $ae^{-\lambda t}$ and $a$
along the flow, in the stable and unstable direction, respectively.
This gives the volume of the single
$3D$ region in the energy shell $v_{1}(t)\sim a^{3}e^{-\lambda t}$.
A typical
distance $d(t)$ between the fettuccine in the overlap regions can be easily estimated as
$d(t)\sim l^{3}a^{-2}e^{-\lambda t}$.
This estimates the number of disconnected regions of the overlap as
\begin{eqnarray}
\label{N1}
N_{1}(t)\sim(l/a)^{3}e^{\lambda t}\,.
\end{eqnarray}
Since $f$ assumes opposite signs in two initial regions, the
distribution $e^{-\hat{L}t}f\partial_{E}\rho_{0}$ consists of two positively and negatively
``charged'' fettuccine. The cancellations result in a signal determined by a typical fluctuation
proportional to $\sqrt{N_{1}}$, and the overlap integral attains a factor $\sqrt{N_{1}}v_{1}$. This
results in the decaying long-time asymptotic of the linear response function $S^{(1)}(t)\sim
e^{-\lambda t/2}$.

We are now in a position to consider the second-order response function $S^{(2)}(t_{1},t_{2})$.
Propagating the observable $f$ in Eq.~(\ref{second-order}) backward in time we interpret the
second-order response as an overlap
\begin{eqnarray}
\label{S-2-FDT} S^{(2)}(t_{1},t_{2})=\partial_{t_{1}}I^{(2)}(t_{1},t_{2})=\partial_{t_{1}}\int
d{\bm\eta}f_{-}\xi^{j}\partial_{j}f_{+}
\end{eqnarray}
of the distributions $f_{-}({\bm\eta})=\exp(\hat{L}t_{2})f$ and
$\xi^{j}\partial_{j}f_{+}({\bm\eta})$ with $f_{+}=\exp(-\hat{L}t_{1})f\partial_{E}\rho_{0}$ and the
vector field $\xi^{j}\partial_{j}=\{f,\cdot\}$. Similar to the case of linear response for $t\gg
1$, the shapes of both distributions $f_{\pm}({\bm\eta})$ become similar to ribbon-like fettuccine,
unchanged along the flow. Since $f_{-}$ results from reverse dynamics, the distributions $f_{+}$
and $f_{-}$ are elongated along the unstable and stable directions of the positive time evolution,
by the factors $e^{\lambda t_{1}}$ and $e^{\lambda t_{2}}$, respectively.
Each of distributions $f_{-}$ and $f_{+}$ consists of two
positively and negatively ``charged'' fettuccine originating from two
oppositely signed separate regions of dipole moment distribution $f$.
Since the vector field ${\bm\xi}$ introduced
earlier is zero outside the support of $f$, the integration in Eq.~(\ref{S-2-FDT}) is restricted to
the overlap of three distributions: $f$ and $f_{\pm}$. The overlap of each of distributions $f_{+}$
and $f_{-}$ with $f$ is represented by $N_{1}(t_{1})$ and $N_{1}(t_{2})$ disconnected fettuccine
pieces as found in Eq. (\ref{N1}).
Therefore the overlap of all
three distributions is represented by $N_{2}(t_{1},t_{2})\sim
N_{1}(t_{1})N_{1}(t_{2})\sim(l/a)^{6}e^{-\lambda(t_{1}+t_{2})}$ disconnected regions. The linear
dimensions along the isoenergetic shell of each disconnected region are estimated as $ae^{-\lambda t_{1}}$,
$ae^{-\lambda t_{2}}$, and $a$, which corresponds to the stable, unstable, and flow
directions, respectively. Therefore the volume of each disconnected region is $v_{2}(t)\sim
a^{3}e^{-\lambda(t_{1}+t_{2})}$.

If instead of the second-order response function, we were dealing with a three-point correlation
function the rest of the story would be straightforward. A three-point correlation function can be
represented in a form of the overlap integral $I^{(2)}(t_{1},t_{2})$ with the differential operator
$\xi^{j}\partial_{j}=\{f,\cdot\}$ replaced by the operator of multiplying by $f({\bm\eta})$.
In full analogy with the linear response case, the typical value of the three-point correlation
results from cancellations of oppositely signed contributions of similar magnitudes and therefore
attains a factor
$\sqrt{N_{2}}v_{2}\sim e^{-\lambda(t_{1}+{t_{2}})/2}$ that describes
the long-time decaying behavior of the correlation function.
The situation with the nonlinear response function is apparently more complicated since generally
the vector field ${\bm\xi}$ has a component along the stable direction. The derivative
$\xi^{j}\partial_{j}$ of a sharp feature in $f_{+}$ along the stable direction can create
exponentially large $e^{\lambda t_{1}}$ factors. This is the Liouville space signature of the
exponentially growing components of the stability matrix, which affects the response starting with
second order due to FDT \cite{MKC96,DellagoMukamel03}. The exponentially growing components of the
stability matrix may seem to be a reason for an exponential divergence of the nonlinear response,
since interaction with the driving field can be considered as a kick leading to an infinitesimal
deviation that grows exponentially with time. This would actually happen if the initial
distribution was $\delta$-functional concentrated at some point in phase space, and the signal
was measured by a deviation of the perturbed trajectory from its unperturbed counterpart. However,
the dipole $f$ that describes the system coupling to the driving field is represented by a smooth
function. As shown below due to this smoothness the exponentially diverging terms cancel out
completely. This demonstrates that response in chaotic systems should be treated using the
Liouville (distribution-based) representation of classical dynamics, while its dual trajectory-based
counterpart may lead to misleading naive picture.

To prove the harmlessness of the derivative in the second-order response function,
we decompose the vector field
${\bm\xi}=\bm{\xi}_{E}+\bm{\xi}_{0}+\bm{\xi}_{+}+\bm{\xi}_{-}$ into the direction along the energy,
flow, unstable, and stable components. This leads to a natural decomposition of the overlap
integral, according to Eq.~(\ref{S-2-FDT}):
\begin{eqnarray}
\label{overlap-decompose} I^{(2)}=I^{(2)}_{E}+I^{(2)}_{0}+I^{(2)}_{+}+I^{(2)}_{-}\,.
\end{eqnarray}
The term $I^{(2)}_{+}$ involves a derivative along the stable direction that provides an additional
exponentially small $e^{-\lambda t_{1}}$ factor, since the distribution $f_{+}$ is elongated along
the unstable direction. The dangerous term that involves derivatives along the sharp feature is
represented by $I^{(2)}_{-}$. The aforementioned cancellation
can be seen after the integration
by parts:
\begin{eqnarray}
\label{cancel-by-parts} I_{-}^{(2)}=-\int d{\bm\eta}{\rm div}{\bm\xi}_{-}f_{-}f_{+}-\int
d{\bm\eta}f_{+}\xi_{-}^{j}\partial_{j}f_{-}\,.
\end{eqnarray}
The first term in Eq.~(\ref{cancel-by-parts}) includes
time-independent distribution
${\rm div}{\bm\xi}_{-}$. The second term contains the derivative in the stable direction of the
distribution $f_{-}$.
Therefore, it acquires an additional exponentially small factor
$e^{-\lambda t_{2}}$ and becomes negligible at long times. Finally, we have at long times
\begin{eqnarray}
\label{qual-long-time} I^{(2)}\approx \int d{\eta}f_{-}\left(-{\rm
div}{\bm\xi}_{-}+\xi_{0}^{j}\partial_{j}+\xi_{E}^{j}\partial_{j}\right)f_{+}\,.
\end{eqnarray}
All terms in Eq.~(\ref{qual-long-time}) do not have derivatives in the stable and unstable
directions that provide exponentially growing or decaying factors. Therefore, according to the
arguments presented above, the second-order response function has a long-time asymptotic
$S^{(2)}(t_{1},t_{2})\sim e^{-\lambda(t_{1}+t_{2})/2}$ similar to the three-point correlation
function. Stated differently, the exponentially growing components of the stability matrices that
in principle enter the expressions for nonlinear response functions
\cite{MKC96,DellagoMukamel03}, do not affect the long-time behavior of the nonlinear response, due
to the smoothness of the dipole function $f({\bm\eta})$ that describes the system coupling to the
driving field.

The qualitative arguments developed in this subsections for the lowest-dimensional case $n=2$ can
be extended to a general case $n\ge 2$ in a straightforward way. This leads to the same asymptotic
expressions for magnitudes of response functions
$S^{(1)}(t)\sim e^{-\lambda t/2}$, and $S^{(2)}(t_{1},t_{2})\sim
e^{-\lambda(t_{1}+t_{2})/2}$, where $\lambda$ should be interpreted in the sense of
Eq.~(\ref{Lyapunov-sum}) as the sum of positive Lyapunov exponents.

\subsection{Effective negative curvature of configuration space}

In subsection \ref{subsection:hyperbolic} we have established a qualitative picture of response in
classical systems with hyperbolic
and mixing
dynamics. In the forthcoming sections we support our
qualitative arguments by performing explicit analytical calculations of the first- and second-order
response functions in a model system of a free particle moving along a Riemann surface of constant
negative curvature. Due to strong dynamical symmetry in this system (see section
\ref{section:free-partcile-2D} for the details) the response functions can be calculated
explicitly. In this subsection we present a rationale why this simple model system can be viewed as
a representative example that bears the basic qualitative features of chaotic dynamics in a
molecule or collection of molecules for high enough energies.

First of all classical motion of a molecular system can be represented as classical dynamics of a
multidimensional particle moving in a potential $U({\bm r})$, where ${\bm r}$ stands for a complete
set of nuclear coordinates. Trajectories of a particle with energy $E$ in arbitrary potential
$U({\bm r})$ are known to be the same as for a free motion in a curved space with the metric
$g_{ik}=(1-U({\bm r})/E)\delta_{ik}$ \cite{Arnold, Gutzwiller}. Although the potential generates
nonuniform motion along the trajectories, the dominant feature of the dynamics is the exponentially
growing separation between the trajectories in the regions where the metric curvature is negative.
The regions of the configuration space that correspond to negative curvature work as defocusing
lenses, causing instability, i.e. divergence of close trajectories. For example, close trajectories
diverge every time they approach the boundary of the classically inaccessible island or pass a
region of a potential local maximum that belongs to the accessible region.
Except for occasional symmetric integrable cases,
passing through the
stable regions with positive curvature cannot compensate for the instability, since stability
reflects oscillatory, rather than converging features in the dynamics of the trajectory deviation.
In particular, existence of unstable regions combined with ergodicity ensures exponential
divergence of trajectories over long enough times.

\begin{figure}[h]
\centerline{
\includegraphics[width=2.2in]{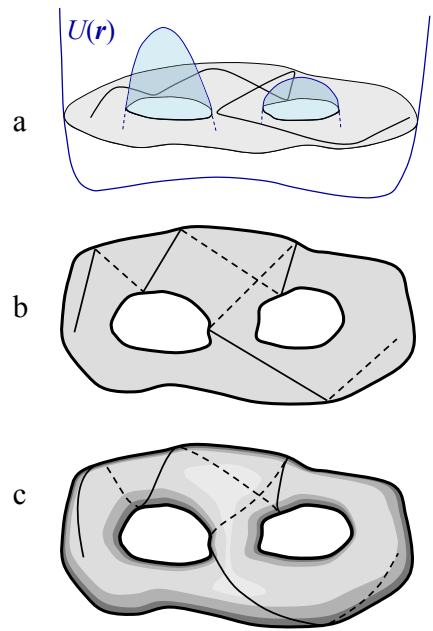}
}
\caption{From the motion in a potential to the motion on compact surface. (a) For fixed particle
energy $E$, a potential with several maxima defines the multiply connected classically allowed
region (b)
Idealization of the hard-wall potential: Trajectories are confined within the original classically
accessible $2D$ configuration space reflecting on its boundaries. Reflections can be viewed as
transitions to the antipode surface component glued to the original one along the boundaries. (c)
Trajectories on the deformed smooth version of the resulting compact surface are shown as solid and
dashed lines, which correspond to the original and antipode components, respectively.
\label{islands} }
\end{figure}

In addition, when the motion is finite, the accessible part of the configuration space at a given
energy can be multiply connected. In the simplest $n=2$ case of two coordinates the motion occurs
inside a disk-like region punctured by $g$ forbidden islands (see Fig. \ref{islands}). Boundaries
of the accessible area are represented by curved lines where the potential energy $U({\bm r})$
coincides with the total energy $E$. Some fraction of trajectories approaches the boundaries so
close that this can be qualified as reflection. Utilizing the original argument of Sinai
\cite{Sinai} reflection can be interpreted as continuing motion on the antipode replica of the
accessible region glued to its original
via the boundaries (see Ref.
\onlinecite{Arnold}). The resulting compact surface has topology of a sphere with $g$ handles
(Riemann surface of genus $g$).
According to the Gauss-Bonnet theorem, in the $g>1$ case the average Gaussian curvature is negative, which
implies regions of instability and results in unstable (hyperbolic) dynamics.

We also note that loose distinction between reflected and deflected trajectories that approach the
boundaries should not be a matter of concern since classical dynamics constitutes an approximation
for a quantum mechanical problem. The approximation is not valid near the boundaries of the
classically accessible region, where the quantum uncertainty takes over. More formally, one can
consider a trajectory reflected (or, equivalently, continued on the antipode replica of the
accessible region) if it penetrates the near-boundary region where quantum effects become
important.
The classical picture of reflection becomes exact for the hard wall potential concentrated on the
boundaries, e.g.  for Sinai billiards \cite{Sinai} and Lorentz gas.

Although in a generic system the effective curvature depends on position ${\bm r}$ in configuration
space, we will rationalize the described above qualitative picture of response in a chaotic system
by calculating the linear and second-order classical response functions for free motion in a
Riemann surface $M^{2}$ of constant negative (Gaussian) curvature. This model that allows an exact
solution has been serving as a prototype of classical chaos, as well as an example of semiclassical
quantization \cite{Arnold, Gutzwiller, BalazsVoros}.

\section{Geometry and dynamics on a Riemann surface with constant curvature}
\label{section:free-partcile-2D}

\subsection{Free particle on a Riemann surface}
In this section we describe classical dynamics of a free particle moving along a Riemann surface
$M^{2}$ of constant negative curvature (Gaussian curvature). This will be done by making use of
strong dynamical symmetry. This system is one of the best studied models of chaotic dynamics.
Although detailed reviews of its symmetry exist in the literature (see e.g. Ref.
\onlinecite{BalazsVoros}), we summarize the necessary results for the sake of completeness and further
extension to the case of Langevin dynamics. Noisy dynamics associated with geodesic flows on
surfaces with constant negative curvature is considered in detail elsewhere \cite{CMResonances}.

The Lagrangian of a free particle contains only kinetic energy: $L=mg_{ik}\dot r^i \dot r^k/2$. The
corresponding unperturbed classical Hamiltonian
\begin{eqnarray}
\label{Hamiltonian}
H({\bm x},\zeta)=\frac{1}{2m} g^{ik}p_i p_k=\frac{\zeta^{2}}{2m}\,.
\end{eqnarray}
does not depend on ${\bm x}$ if expressed in terms of the absolute value of the momentum $\zeta$.
Here we introduced covariant and contravariant metric tensors $g_{ik}$ and $g^{ik}$.
The curvature (also referred to as the Gaussian curvature)
of configuration space $M^2$ is expressed in terms of derivatives
of the metric tensor.

Since unperturbed classical dynamics conserves energy, the value of $\zeta$ is preserved, and
evolution actually occurs in the reduced phase space (energy shell) $M^{3}$. If we are not
interested in a trivial case of zero energy, distributions can be thought of as functions
$\rho({\bm x},\zeta)$, with ${\bm x}\in M^{3}$ and $\zeta>0$ is the absolute value of the particle
momentum. A compact $3$-dimensional smooth manifold $M^{3}$ is the subspace of the phase space that
consists of points with unit length of the momentum vector. Points $\bm x\in M^{3}$ are specified
by two coordinates $\bm r$ and momentum direction angle $\theta$.
The dynamical symmetry for the free-particle dynamics originates from the fact that there is a
smooth action
of the group $G\cong SO(2,1)$ in the reduced phase space of the system, consistent with the
dynamics. The details are presented in Appendix \ref{appendix:symmetry}. The dynamical symmetry has
a simple and clear interpretation in infinitesimal terms, i.e. action of the corresponding Lie
algebra $so(2,1)$, where the algebra generators are implemented as vector fields in $M^{3}$.

As discussed in Section \ref{section:classical-liouville}, the evolution in the phase space is
determined by equation $\dot{\bm\eta}=\hat L \bm\eta$. In this equation $\hat L \bm\eta$ is a
vector field that belongs to the tangent space of phase space point $\bm\eta$. Hereafter we adopt
an agreement used in differential geometry by identifying a first-order differential operator of
differentiating along the vector field with the vector field itself. Due to energy conservation,
vector field $\hat L$ is tangent to the reduced phase space $M^3$.
We introduce a vector field  $\sigma_{1}$ that determines the geodesic flow and generates the
 translation along trajectories, so that the Liouville operator of free motion is
\begin{eqnarray}
\nonumber
\hat L=\zeta\sigma_1\,.
\end{eqnarray}

Another natural vector field in $M^3$ is $\sigma_{z}$ which corresponds to the generator of the
momentum rotation and keeps the coordinates unchanged. This operator can be represented as
$\sigma_{z}=\partial/\partial\theta$ in terms of a derivative with respect to the particle momentum
direction $\theta$.

Operators $\sigma_1$ and $\sigma_z$ can be easily expressed in terms of $\bm r$ and $\bm p$.
Performing the variables transformation we obtain:
\begin{align}
\nonumber
&
\sigma_1=\frac{\partial\zeta}{\partial p_i}\frac{\partial}{\partial r_i}-
\frac{\partial\zeta}{\partial r_i}\frac{\partial}{\partial p_i}\,,
\quad
\sigma_z=E_{ji}g^{jk}p_k\frac{\partial}{\partial p_i}\,.
\end{align}
Here $E_{ij}$ is the asymmetric tensor, $E_{12}=-E_{21}=(\det g^{ik})^{-1/2}$,
$E_{11}=E_{21}=0$.
Both operators $\sigma_1$ and $\sigma_z$ obviously conserve energy:
$\sigma_1\zeta=\sigma_z\zeta=0$.

Finally, we introduce the third vector field in $M^{3}$ as $\sigma_{2}=[\sigma_{1},\sigma_{z}]$
where the commutator of vector fields is understood as the commutator of the corresponding
differential operators. A simple calculation shows that in the case of constant negative curvature
the vector fields $\sigma_{z}$, $\sigma_{1}$, and $\sigma_{2}$ form the Lie algebra $so(2,1)$ with
respect to the vector field commutator (see Appendix \ref{appendix:poisson} for more details):
\begin{eqnarray}
\label{so(2,1)-commutation}
[\sigma_{1},\sigma_{2}]=\sigma_{z}\,, \;\;\;
[\sigma_{1},\sigma_{z}]=\sigma_{2}\,, \;\;\;[\sigma_{2},\sigma_{z}]=-\sigma_{1}\,.
\end{eqnarray}
The group action of $SO(2,1)$ in the reduced phase space $M^{3}$ is obtained by integration of the
algebra action.

Any compact surface of constant negative curvature may be represented as a result of factorization
of the hyperbolic plane with respect to translations that constitute a finitely generated infinite
discrete subgroup of $SO(2,1)$ (see Appendix \ref{appendix:symmetry}). The entire hyperbolic plane
possesses the complete $SO(2,1)$ symmetry, which makes the infinite motion there integrable due to
the presence of the integral of motion similar to the angular momentum in the case of $SO(3)$. The
factorization destroys global $SO(2,1)$ symmetry and folds back the trajectories, which renders the
dynamics strongly chaotic.

The Poisson bracket can be computed in a standard way, making use of the fact that locally it coincides
with the Poisson bracket for a free particle moving in the entire hyperbolic plane $H$
(see Appendix \ref{appendix:poisson}).
The canonical Poisson bracket may be rewritten in terms of the linear differential operators
introduced above
acting on two functions $f({\bm x},\zeta)$ and $g({\bm x},\zeta)$ as
\begin{eqnarray}
\label{Poisson-bracket-2}
&&
\{f,g\}=
\\
\nonumber
&&
\frac{\partial f}{\partial\zeta}(\sigma_{1}g)-(\sigma_{1}f)\frac{\partial g}{\partial\zeta}
+\frac{1}{\zeta}\left((\sigma_{2}f)(\sigma_{z}g)-(\sigma_{z}f)(\sigma_{2}g)\right)\,.
\end{eqnarray}
The Poisson bracket in Eq.~(\ref{Poisson-bracket-2}) is expressed as a bilinear form of generators
of a Lie algebra with constant coefficients. This is is another important manifestation of the
dynamical symmetry in the problem of free motion on the surface of constant curvature.

The form of the Poisson bracket as well as the commutation relations suggest that the Lyapunov
exponent is equal to $\lambda=\sqrt{-K}$. This can be clearly seen from the
Jacobi equation $\ddot y+Ky=0$ for the magnitude of the normal component of the deviation from a
given geodesic\cite{Arnold}. The stable and unstable directions are given by linear combinations of
$\sigma_2$ and $\sigma_z$. The displacements along the flow $\sigma_1$ are conserved. The deviation
of the trajectory at $t=0$ can be written as $\delta\bm\eta(0)=\sigma(\bm\eta)$, where $\sigma$ is
a first order differential operator corresponding to a certain direction in the phase space. The
evolution of the phase point $\bm\eta+\delta\bm\eta(0)$ is given by $e^{\sigma_1
t}(\bm\eta+\sigma\bm\eta)=\bm\eta(t)+e^{\sigma_1 t}\sigma e^{-\sigma_1 t}\bm\eta(t)$. We see that
after time $t$ the deviation from the phase point $\bm\eta(t)$ is determined by the vector field
$\sigma(t)=e^{\sigma_1 t}\sigma e^{-\sigma_1 t}$.
Commutation relations (\ref{so(2,1)-commutation}) allow to find this Heisenberg representation
of any operator decomposed over the basis elements of $so(2,1)$.
In particular, we find that
\begin{equation}
e^{\sigma_1 t}(\sigma_2\pm\sigma_z)e^{-\sigma_1 t}= e^{\pm t}(\sigma_2\pm\sigma_z)\,,
\end{equation}
and thus conclude that the local stable and unstable directions are given
by $\sigma_2-\sigma_z$ and $\sigma_2+\sigma_z$, respectively.
The fact that the form of these vector fields is conserved by the dynamics is an important
manifestation of the dynamical symmetry.

Dynamical symmetry also leads to the following general relation:
\begin{eqnarray}
\label{invar-int} \int_{M^{3}}d{\bm x}\sigma_{l}f({\bm x})=0
\end{eqnarray}
for $l=1$, $2$ or $z$ that is valid for any smooth function $f({\bm x})$ where $d{\bm x}$ in
$M^{3}$ is an invariant integration measure with respect to the $SO(2,1)$ action (see Appendix
\ref{appendix:symmetry} for details). Eq.~(\ref{invar-int}) implies an integration-by-part rule
\begin{eqnarray}
\label{integr-by-parts} \int_{M^{3}}d{\bm x}f({\bm x})\sigma_{l}g({\bm x})=-\int_{M^{3}}d{\bm
x}\left(\sigma_{l}f({\bm x})\right)g({\bm x})\,,
\end{eqnarray}
that will be an important ingredient of our analytical calculations.

In conclusion we emphasize that the dynamical symmetry with respect to the action of the group
$G\cong SO(2,1)$ does not mean symmetry in a usual sense, i.e. that the system dynamics commutes
with the group action, but rather reflects the fact that the vector field $\hat L$ that determines
the classical dynamics is represented by an element of the corresponding Lie algebra $so(2,1)$.
This allows to apply representation theory as described below.

\subsection{Decomposition of the free-particle dynamics using irreducible representations}

The smooth action of $G$ in $M^{3}$ can be interpreted as that the space ${\cal H}$ of smooth
functions in $M^{3}$ constitutes a representation of $G$, which turns out to be a unitary
representation (see Refs. \onlinecite{Kirillov,Lang,Williams} and Appendix \ref{appendix:symmetry}),
and therefore can be decomposed into a
direct sum of irreducible representations of $G$. The spectrum ${\rm
Spec}_{0}(M^{2})\subset\hat{G}$ of the Riemann surface is defined as a discrete subset of the space
$\hat{G}$ of irreducible unitary representations of the principal series that participate in the
decomposition of functions in $M^{3}$ into irreducible representations:
\begin{eqnarray}
\label{decompose-distribution}
\rho({\bm x},\zeta)=g^{(0)}(\zeta)+
\sum_{s\in{\rm Spec}_{0}(M^{2})}g_{s}({\bm x},\zeta)\,.
\end{eqnarray}
The spectrum ${\rm Spec}_{0}(M^{2})$ consists of imaginary numbers $s$ that characterize
representations of the principal series. The effects of the inclusion of other (complementary and
discrete) series are discussed after their review in Appendices \ref{appendix:symmetry} and
\ref{appendix:irreducible-SO(2,1)}.
The evolution in the space of distributions is decomposed into a set of uncoupled evolutions that
correspond to relevant irreducible representations. The component dynamics is determined by the
reduced Liouville operators:
\begin{eqnarray}
\label{irred-Liouv-eq}
\frac{\partial g_{s}({\bm x},\zeta;t)}{\partial t}=
-\zeta\hat{L}(s) g_{s}({\bm x},\zeta;t)\,.
\end{eqnarray}
where $\hat L(s)$ corresponds to an irreducible representation labeled by $s$.
The distribution components can be further decomposed as
\begin{eqnarray}
\label{decomp-ang-mom}
g_{s}({\bm x},\zeta)=
\sum_{k=-\infty}^{\infty}\rho_{s,k}(\zeta)\psi_{k}({\bm x};s)\,,
\end{eqnarray}
using the eigenstates of the angular momentum operator $\sigma_z$. These satisfy the following
properties:
\begin{align}
\label{sigma-act-irred0}
&
\sigma_{z}\psi_{k}({\bm x};s)=ik\psi_{k}({\bm x};s)\,,
\\
\nonumber & \sigma_{\pm}\psi_{k}({\bm x};s)=\left(\pm k+\frac{1}{2}-s\right)\psi_{k\pm 1}({\bm
x};s) \,,
\end{align}
where we introduced the raising and lowering operators $\sigma_{\pm}=\sigma_{1}\pm i\sigma_{2}$
that are anti-Hermitian conjugated, i.e. $\sigma_{+}^{\dagger}=-\sigma_{-}$.

The description of unitary representations of $G$, reviewed in detail in Appendix
\ref{appendix:irreducible-SO(2,1)} (see also Refs. \onlinecite{Lang,Williams}), allows to represent
the eigenstates $\psi_{k}({\bm x};s)$ by the functions on the circle $\Psi_{k}(u)$ for any given
$s\in{\rm Spec}(M^{2})$. In the case of imaginary $s$ (principal series), when $\Psi_{m}(u)$
constitutes an orthogonal normalized set with the natural scalar product, this leads to normalized
functions $\psi_{k}({\bm x};s)$.
Identification of the functions $\psi_{k}(\bm x)$ with $\Psi_{k}(u)$ establishes an isomorphism
between two irreducible representations, the first being ${\cal H}_{s}$ that participates in the
decomposition of Eq.~(\ref{decompose-functions}), the second being its standard representation in
function in a circle described in Appendix \ref{appendix:irreducible-SO(2,1)}. According to the
Shur lemma (see, e.g. \cite{Kirillov}) the identification (isomorphism) is determined up to a
factor. Its absolute value can be fixed by requiring that the function $\psi_{0}(\bm x;s)$,
identified with $\Psi_{0}(u)=1$ is normalized. To fix also its phase we note that $\psi_{0}(\bm
x;s)$ does not depend on $\theta$
and turns out to be an eigenfunction of the Laplacian operator $\nabla^{2}$ in $M^{2}$ as described
by Eq.~(\ref{Casimir}). Since the Laplacian is a real operator, the eigenfunction $\psi_{0}({\bm
x};s)$ can be chosen to be real. Hereafter, we implement an agreement that $\psi_{0}(\bm x;s)$
identified with $\Psi_{0}(u)=1$ is real. This determines the functions up to a sign, the latter
being of no importance: $\Psi_k(u)=e^{iku}$.

The form of $\sigma_1$ in the angular representation is fixed up to a phase \cite{Lang}:
\begin{eqnarray}
\label{rho-circle}
\hat{L}(s)=\sigma_{1}=\sin u\frac{d}{du}+\frac{1-2s}{2}\cos u\,.
\end{eqnarray}
Thus, the original dynamics of a distribution $g_s(\bm x;s)$ in Eq. (\ref{decomp-ang-mom}) is
mapped onto an effective classical dynamics of a distribution ${\cal
G}(u)=\sum_{k=-\infty}^{\infty}\rho_{s,k}(\zeta)\Psi_{k}(u;s)$ defined on a circle with the
Liouville operator (\ref{rho-circle}). The resulting effective problem is one-dimensional and can
be solved exactly.

The term that describe the interaction with the driving field can be also decomposed in irreducible
representations. The coupling (polarization) $f$ is represented by a function of the particle
position only (a function in $M^{2}$) or, stated differently, a phase-space function that does not
depend on the particle momentum. It can be equivalently interpreted as a function in $M^{3}$
independent of $\theta$, hence $\sigma_{z}f=0$. Thus, $f$ can be expanded as
\begin{eqnarray}
\label{f-expansion}
f=\sum_{s\in{\rm Spec}_{0}(M^{2})}B_{s}\psi_0(\bm x;s)\,.
\end{eqnarray}
The sum in Eq.~(\ref{f-expansion}) runs over the spectrum of the compact surface corresponding to
the principal series representations characterized by imaginary $s$ (see Appendix
\ref{appendix:irreducible-SO(2,1)}). We emphasize that, as noted earlier, the functions
$\psi_{0}({\bm x};s)$, being actually functions of the particle position only, are the
eigenfunctions of the Laplacian operator $\nabla^{2}$ on the Riemann surface. Typically the dipole
$f$ is a slow function of coordinates, and therefore only few terms provide substantial
contributions to the expansion of Eq.~(\ref{f-expansion}). We also note that, although the
Laplacian diagonalization on an arbitrary Riemann surface with constant curvature is a complex
problem, its eigenfunctions are tangible and intuitively simple objects. Combined with the previous
results, the expansion (\ref{f-expansion}) leads to closed analytical expressions for the response
functions of the original problem.

\subsection{Effective dynamics on the circle}
\label{subsection:dynamics-on-circle}

We have identified the dynamical symmetry that allows to map the original chaotic dynamics on $M^2$
onto a tractable dynamical problem on the circle. In this subsection we consider reduced dynamics
that corresponds to a principal series representation labeled by $s$ with ${\rm Im}s>0$. In the
following calculations of the response functions, we will need the expansion of $e^{-\hat L
t}\psi_0(\bm x;s)$ over basis vectors $\psi_k({\bm x};s)$,
\begin{align}
\label{evolution-expansion} e^{-\hat L t}\psi_0(\bm
x;s)=\sum\limits_{k=-\infty}^{+\infty}A_k(t;s)\psi_k(\bm x;s)\,.
\end{align}
In particular, the linear response is expressed in terms of the coefficient $A_0$. These
coefficients that are actually matrix elements of the evolution operator between the angular
harmonics have been calculated earlier \cite{BalazsVoros,RobertsMuz}. The result immediately
follows from the description of irreducible representations using a construction of an induced
representation (see, e.g. Ref. \onlinecite{Lang}). The derivation presented below allows an
extension to the Langevin dynamics, associated with the original classical dynamical problem. The
picture based on the Fokker-Planck equation will be presented in detail elsewhere
\cite{CMResonances}.

The coefficients $A_k(t;s)$ are calculated by implementing the representation on the circle,
introduced above. Since $e^{-\hat L(s) t}\Psi_0(u)$ represents a function obtained as the result of
the evolution operator action on $\Psi_0(u)\equiv 1$, it can be found by solving the dynamical
equation $\partial_t g(t,u)+\hat L g(t,u)=0$. For the principal series representation labeled by
$s$, the equation adopts the following form:
\begin{eqnarray}
\label{equation-on-circle}
\partial_t g(t,u)+\left(\sin u\frac{\partial}{\partial u}+\frac{1-2s}{2}\cos u\right)g(t,u)=0
\,.
\end{eqnarray}
The solution of  the first order partial differential equation supplemented with an initial
condition $g(0,u)=1$ can be found using the method of characteristics:
\begin{align}
\nonumber
&
e^{-\hat L(s) t}\Psi_0(u)=e^{st-t/2}
\left|\cos \frac{u}{2} \right|^{2s-1}
\left(1+ e^{-2t}\tan^2\frac{u}{2}\right)^{s-\frac{1}{2}}
\\
\label{solution}
&
=
\left(\cosh t + \sinh t\cos u\right)^{s-\frac{1}{2}}
\,.
\end{align}
At long times $t>0$ the solution is concentrated near the stable stationary point $u=\pi$. This
reflects the collapse of the reduced phase space distribution function along the stable direction.
Indeed, the width of the region where $e^{-\hat L(s) t}\Psi_0(u)$ is not exponentially small
vanishes as $\propto e^{-t}$, according to the fact that for the chaotic dynamics on $M^2$ with
Gaussian curvature $K$ the Lyapunov exponent is equal to $\sqrt{-K}$.

The expansion coefficients in Eq.~(\ref{evolution-expansion}) are given by
\begin{eqnarray}
\label{Fourier-coefficient1}
A_k(\zeta t;s)=\int\frac{du}{2\pi}\Psi_k^*(u) e^{-\zeta\hat L(s) t}\Psi_0(u)\,.
\end{eqnarray}
The integral can be reduced to a standard integral representation of the Gauss hypergeometric
function $\phantom{z}_2F_1(a,b,c,z)$ defined as a series
\begin{eqnarray}
\label{hypergeometric-series}
\phantom{z}_2F_1(a,b,c,z)=\sum\limits_{n=0}^\infty \frac{\Gamma(a+n)\Gamma(b+n)\Gamma(c)}
{\Gamma(a)\Gamma(b)\Gamma(c+n) n!}z^n\,,
\end{eqnarray}
$\Gamma(z)$ being the gamma function \cite{specfunctions}. An exact expression for $k\ge 0$ reads:
\begin{align}
\label{Fourier-coefficient2}
&
A_k(t;s)=
(-1)^k\frac{\Gamma\left(k+\frac{1}{2}-s\right)}{k!\Gamma\left(\frac{1}{2}-s\right)}
\left(\tanh\frac{t}{2}\right)^k\times
\\
&
\nonumber
\phantom{|}_2F_1\left(\frac{1}{2}+s,\frac{1}{2}-s,k+1,-\sinh^2\frac{t}{2}\right)\,.
\end{align}
Various representations of these coefficients using other special functions have been derived in
the context of two-point correlations \cite{BalazsVoros,RobertsMuz}.

Since solution $g$ is even, $g(t,-u)=g(t,u)$, coefficients $A_k$ are symmetric, $A_{-k}=A_k$. The
time reversal property
\begin{align}
\label{time-reversal} g(-t,u)=g(t,u+\pi)\,.
\end{align}
that follows directly from Eq. (\ref{equation-on-circle}), immediately implies $A_k(-t)=(-1)^k
A_k(t)$. We can represent $A_k(t;s)$ in an alternative form suitable for studying its long-time
behavior \cite{specfunctions}:
\begin{align}
\label{Fourier-coefficient3}
&
A_k(t;s)=\frac{2(-1)^k\left(1-e^{-2t}\right)^k \Gamma(k+\frac{1}{2}-s)}
{\sqrt{\pi}\Gamma(\frac{1}{2}-s)}e^{-t/2}\times
\\
&
\nonumber
{\rm Re}
\Bigl(
\frac{\Gamma(s)e^{st}}{\Gamma(k+\frac{1}{2}+s)}
\phantom{|}_2F_1\Bigl(k+\frac{1}{2}-s,k+\frac{1}{2},1-s,e^{-2t}\Bigr)
\Bigr).
\end{align}
Its expansion in $e^{-2t}$ using Eq. (\ref{hypergeometric-series}) corresponds to the spectral
decomposition of the regularized evolution operator for the irreducible representation labeled by
$s$. The properly interpreted eigenvalues of the Liouville operator are known as Ruelle-Pollicott
resonances for chaotic systems. They can be obtained using a physical approach. Adding weak
Langevin noise results in adding an infinitesimal second-order differential diffusion operator to
the classical Liouville operator. This results in a Fokker-Planck operator
 $\hat{\cal L}=-\kappa\nabla^{2}+\hat{L}$
that represents a regularized version of the Liouville operator
\cite{Gaspardbook}. The resonances are more often treated as mathematical objects in a form of
generalized eigenfunctions in a properly chosen rigged Hilbert space (see e.g. Ref.
\onlinecite{RobertsMuz}). The physical approach to spectral decomposition of linear and nonlinear
response functions for a free particle moving on a surface of constant negative curvature has been
developed in Ref. \onlinecite{CMResonances}.

We will be interested in long-time behavior of the response functions and expect that the
asymptotics
\begin{align}
\label{Fourier-coefficient-asympt}
A_k(t;s)\approx \frac{2(-1)^k\Gamma(k+\frac{1}{2}-s)}
{\sqrt{\pi}\Gamma(\frac{1}{2}-s)}e^{-t/2}
{\rm Re}\left(
\frac{\Gamma(s)e^{st}}{\Gamma(k+\frac{1}{2}+s)}
\right)
\end{align}
at $t\to \infty$ are relevant. However, we will see that, since the approximation breaks down for
higher harmonics $k \gtrsim e^{t}$ and does not vanish at $k\to\infty$, it cannot be immediately
utilized for nonlinear response calculations where the result is given by the infinite series over
$k$. In the region $t\gtrsim 1$ and $1\ll k\ll e^{2t}$ one can use an approximation for $A_k(t)$ in
terms of the MacDonald function (modified Bessel function of the second kind):
\begin{align}
A_k(t;s)\approx \frac{2(-1)^k}{\sqrt{\pi}\Gamma(\frac{1}{2}-s)} e^{-t/2}
k^{-s}K_{s}(2ke^{-t})\,.
\end{align}

\section{Calculation of response functions}
\label{section:calculation}

In this section we substantiate the qualitative picture of response presented in Section
\ref{section:qualitative} which is based on general expressions (\ref{first-order},
\ref{second-order}) for a particular model of free motion on a compact surface with constant
negative curvature. Heretofore we have not made any assumptions about the shape of the initial
distribution $\rho_0$ in the general expressions for response functions. In the case of free
motion, the equilibrium phase space density $\rho_0$ depends on the momentum absolute value only.
The fluctuation-dissipation theorem (FDT) relates the linear response function to the correlation
function in a system at thermal equilibrium. To reduce bulky calculations, in both linear and
second-order response functions, we employ an analog of FDT and notice that the action of the
evolution operator on the most inner Poisson bracket can be reduced to the time derivative. It
allows recasting Eqs. (\ref{first-order},\ref{second-order}) in a form:
\begin{align}
\label{first-order1}
&
S^{(1)}(t_1)=\partial_{t_1}\int d{\bm \eta}\,
f({\bm \eta})e^{-\hat L t_1}f({\bm \eta})\frac{\partial\rho_0}{\partial \zeta}\,.
\\
\label{second-order1}
&
S^{(2)}(t_1,t_2)=\partial_{t_1}\!\int \! d{\bm \eta}\,
f({\bm \eta})e^{-\hat L t_2}
\left\{f({\bm \eta}),e^{-\hat L t_1}\!f({\bm \eta})\frac{\partial\rho_0}{\partial\zeta}\right\}.
\end{align}

The general expressions given by Eqs.~(\ref{first-order1}) and (\ref{second-order1}) apply to any
equilibrium distribution, including the canonical $\rho_0\propto e^{-\beta E}$, microcanonical
$\rho_0\propto \delta(E-E_{0})$, and cumulative microcanonical $\partial_{E}\rho_0\propto
\delta(E-E_{0})$ distributions. To obtain the final expressions in the most transparent form we
will use the cumulative microcanonical distribution with the cut-off energy set to $E_{0}=1/2$.

We are able to approach an apparently intractable problem of calculating the response in a chaotic
system because of strong dynamical symmetry present.
The preliminary steps of our calculation have been described in the preceding section. Dynamical
symmetry leads to decomposition of the original dynamical problem into a set of uncoupled
evolutions that correspond to irreducible representations of the dynamical symmetry group
$SO(2,1)$. The dipole moment $f$ is expanded over momentum-independent basis functions $\psi_0(\bm
x;s)$ according to Eq. (\ref{f-expansion}), where $s$ labels relevant irreducible representations.
The relevant matrix elements $A_k(t;s)$ of the evolution (Perron-Frobenius) operator are given by
Eq.~(\ref{Fourier-coefficient2}). The coefficients $A_k(t;s)$ constitute the dynamical part in the
calculation of the response functions. The geometrical part of the calculation is represented by
the integrals over the reduced phase space, hereafter referred to as (geometrical) matrix elements.
The geometrical part turns out to be trivial in the linear response case while the calculations for
the second-order response are more involved.
Computation of the matrix elements involved in the nonlinear response is one of the main technical
results of this manuscript. The second-order response function will be obtained in a form of an
infinite series. A nontrivial character of the series convergence reflects a nontrivial way how the
exponentially growing terms cancel out in the nonlinear response, as described in Section
\ref{subsection:hyperbolic}.

In this section for the sake of simplicity the calculations are performed in the case when the
dipole moment $f$ has a single component $f=\psi_0(\bm x;s)$ in the expansion over irreducible
representations (\ref{f-expansion}). Stated differently, the dipole function $f$ is represented by
a single eigenmode of the Laplace operator. Some details for the case of two components,
$f=B_{s_1}\psi_0(\bm x;s_1)+B_{s_2}\psi_0(\bm x;s_2)$ (superposition of two Laplacian eigenmodes),
are presented in Appendix \ref{appendix:matrix-elements}. Numerical results for the second-order
response function in this simplest case that involves different resonances are given in Section
\ref{section:numerical}.

\subsection{Linear response}
\label{subsection:linear-response}

Since the basis functions $\psi_k(\bm x;s)$ with different $k$ are mutually orthogonal, the linear
response function is determined by the dynamical part alone and can be represented in a form
\begin{align}
\label{linear-response-result-1}
S^{(1)}(t)=
\frac{\partial}{\partial t}
\int\limits_0^\infty d\zeta\, A_0(\zeta t;s)\frac{\partial \rho_0}{\partial \zeta}\,,
\end{align}
with
\begin{align}
&
A_0(t;s)=
\frac{2}
{\sqrt{\pi}}e^{-t/2}
\times
\\
&
\nonumber
{\rm Re}\left(
\frac{\Gamma(-s)e^{-st}}{\Gamma(\frac{1}{2}-s)}
\phantom{z}_2F_1\left(\frac{1}{2}+s,\frac{1}{2},1+s,e^{-2t}\right)
\right).
\end{align}
according to Eq.~(\ref{Fourier-coefficient3}). For large $t$ the linear response function shows
oscillatory decay as $e^{-(1/2\pm s)t}$. The expansion in powers of $e^{-2t}$
can be interpreted as the expansion over Ruelle-Pollicott resonances. The resonances can be found
as the eigenvalues of the regularized Liouville operator \cite{CMResonances}. Only even resonances
$\omega_{2k}=2k+1/2 \pm s$ contribute to the response function. The expansion is a convergent
series over $e^{-2t_1}$ if $e^{-2t_1}<1$, i.e. $t_1>0$.

In a general case the coupling $f(\bm x)$ can be decomposed in irreducible representations labeled
by $s$ [ see Eq.~(\ref{f-expansion})], and the linear response function becomes a linear
combination of contributions (\ref{linear-response-result-1}) with the coefficients $B_s^2$. Due to
orthogonality of the basis functions, there is no coupling between different representations.

\subsection{Second-order response: matrix elements}
\label{section:matrix-elements}

The second-order response function for a free particle on a Riemann surface with constant negative
curvature is obtained by substituting the specific expression for the Poisson bracket
[Eq.~(\ref{Poisson-bracket-2})] into the generic expression given by Eq.~({\ref{second-order1}}).
We further simplify the calculation by making use of the evolution operator unitarity $\left(
e^{-\hat L t_2} \right)^\dag=e^{\hat L t_2}$ with respect to the natural scalar defined by Eq.
(\ref{orth-cond-M3}).
Performing the integration over $\zeta$ by parts and making use of
$\partial\rho_0/\partial\zeta|_{\zeta=0,\infty}=0$ results in:
\begin{align}
\label{S2}
&
S^{(2)}(t_1,t_2)=
\frac{\partial}{\partial t_1}
\int d{\bm x}\,d\zeta\,
\times
\\
\nonumber
&
\left(
\zeta t_2\left(\sigma_1 e^{\sigma_1\zeta t_2}f\right)^*
(\sigma_1f)(e^{-\sigma_1\zeta t_1}f)
\right.
\\
\nonumber
&
+\left(e^{\sigma_1\zeta t_2}f\right)^*
(\sigma_1f)(e^{-\sigma_1\zeta t_1}f)
\\
\nonumber
&
\left.
+
\left( e^{\sigma_1\zeta t_2}f\right)^*
(\sigma_2f)(\sigma_z e^{-\sigma_1\zeta t_1}f)
\right)
\frac{\partial\rho_0}{\partial\zeta}\,.
\end{align}

The transformation to Eq.~(\ref{S2}) is an important step towards expressing the result in terms of
the quantities calculated in the previous sections. This allows to avoid propagating the vector
fields $\sigma_2$ and $\sigma_z$, and instead deal only with the action of the evolution operators
$e^{\hat L t_2}$ and $e^{-\hat{L}t_1}$ on the dipole moment $f$. The latter is decomposed according
to Eq.~(\ref{f-expansion}) in zero harmonics $\psi_0(\bm x;s)$ of irreducible representations
labeled by $s$, i.e. the Laplacian eigenmodes. Then we use the expansion
(\ref{evolution-expansion}) to express $e^{-\hat{L}t}\psi_0(\bm x;s)$ in terms of the angular
harmonics $\psi_k(\bm x;s)$ which constitutes a basis set in the representation.

Now we approach the most challenging part of the calculation represented by the integration over
the reduced phase space. Generally, according to the decomposition of $f$, the integration may
involve functions from different irreducible representations. In this section we consider the
simplest case $f=\psi_0(\bm x;s)$ of the dipole function represented by a single eigenmode of the
Laplacian. A more complicated case of two representations is treated in Appendix
\ref{appendix:matrix-elements}.

One needs to find a way to calculate the integrals of triple products like $\psi^*_n(\bm
x;s)(\sigma_{1,2}\psi_0(\bm x;s))\psi_m(\bm x;s)$ over the reduced phase space $M^3$. This matrix
element in the second-order response function is essentially less trivial part than its counterpart
in the linear response function, where the integral of the double product of $\psi_n(\bm x;s)$ has
been done based on orthogonality and normalization of the functions. The problem arises from the
fact that integrals of relevant triple products over the reduced phase space do not have natural
representation in terms of the effective system on the circle. Nevertheless, we will show that this
geometrical part of the second-order response can be worked out using relatively simple tools, and
can be expressed in terms of few quantities related to the Laplacian eigenmodes.

The integration in reduced phase space $\bm x=(\bm r,\theta)$ is performed over the particle
coordinates $\bm r \in M^2$ and the locally defined angle $\theta$ that represents the $2D$
momentum direction. The local (for fixed $\bm r$) dependence of $\psi_n(\bm x;s)$ on $\theta$ is
given by $\psi_n(\bm x;s)\propto e^{in\theta}$ due to the definition $\partial_\theta\psi_n(\bm
x;s)=in\psi_n(\bm x;s)$ of the basis functions. The operators $\sigma_1$ and $\sigma_2$ applied to
$\psi_0(\bm x;s)$ in the middle of the integrand are expressed via the ladder operators
$\sigma_\pm$. Therefore, integration over $\theta$ in Eq. (\ref{S2}) results in zero unless the
integrand is locally independent of $\theta$, i.e.
\begin{eqnarray}
\label{triple-product}
\int_{M^{3}}d{\bm x}\left(\psi_{n}({\bm x};s)\right)^{*}
(\sigma_\pm\psi_{0}({\bm x};s))
\psi_{m}({\bm x};s)\propto \delta_{n,\,m\pm 1}\,.
\end{eqnarray}
We now introduce the matrix element $a_k$  and $b_k$ as
\begin{eqnarray}
\label{am}
a_k=
\int_{M^{3}}d{\bm x}\psi_{k}^{*}({\bm x};s)
\psi_{0}({\bm x};s)
\psi_{k}({\bm x};s)\,,
\\
\label{b-k}
b_k= 2 \int_{M^{3}}d{\bm x}\psi_{k}^{*}({\bm x};s) (\sigma_+\psi_{0}({\bm x};s))
\psi_{k-1}({\bm x};s)\,.
\end{eqnarray}
As established in Section \ref{section:free-partcile-2D}, the differential operators $\sigma_\pm$
can be moved from one part of the integrand to another one, and thus one can integrate by parts
[Eq.~(\ref{integr-by-parts})]. Integrating by parts in Eq.~(\ref{b-k}) and making use of the
properties (\ref{sigma-act-irred0}) of $\sigma$ operators we derive the following implicit
recurrence relations:
\begin{align}
\label{recur-deriv1}
&
b_{k+1}=
2
\int d\bm x\,\psi_{k+1}^*(\sigma_+\psi_0)\psi_k=
\\
&
\nonumber
-2\int d\bm x\,(\sigma_-\psi_{k+1})^*\psi_0\psi_k-2\int dx\,\psi_{k+1}^*\psi_0\sigma_+\psi_k=
\\
&
\nonumber
(2k+1-2s)(a_k-a_{k+1})\,,
\end{align}
and
\begin{align}
\label{recur-deriv2}
&
b_{k+1}=
2
\int d\bm x\,\psi_{k+1}^*(\sigma_+\psi_0)\psi_k=
\\
&
\nonumber
\frac{4}{2k+1+2s}\int d\bm x\,(\sigma_+\psi_k)^*(\sigma_+\psi_0)\psi_k=
\\
&
\nonumber
-\frac{4}{2k+1+2s}
\int d\bm x\,\psi_k^*
\bigl((\sigma_-\sigma_+\psi_0)\psi_k+(\sigma_+\psi_0)\sigma_-\psi_k\bigr)=
\\
&
\nonumber
\frac{1-4s^2}{2k+1+2s}a_k+\frac{2k-1+2s}{2k+1+2s}b_k\,.
\end{align}
We further make use of Eq. (\ref{recur-deriv1}) to express $a_{k+1}$ via $a_k$ and $b_{k+1}$, and
applies Eq. (\ref{recur-deriv2}) to find $b_{k+1}$ in terms of $a_k$ and $b_k$. This allows to
recast the recurrence relations between the matrix elements in an explicit form:
\begin{align}
\label{recurrent-ab} & a_{k+1}=\frac{4k(k+1)}{(2k+1)^2-4s^2}a_k
-\frac{2k-1+2s}{(2k+1)^2-4s^2}b_k,
\\
&
\nonumber
b_{k+1}=\frac{1-4s^2}{2k+1+2s}a_k+\frac{2k-1+2s}{2k+1+2s}b_k\,.
\end{align}
The matrix elements $b_k$ can be excluded to connect three adjacent matrix elements $a_{k-1}$,
$a_k$ and $a_{k+1}$:
\begin{eqnarray}
\label{recurrent}
a_{k+1}=\frac{8k^2+1-4s^2}{(2k+1)^2-4s^2}a_k-\frac{(2k-1)^2-4s^2}{(2k+1)^2-4s^2}a_{k-1}\,.
\end{eqnarray}

As mentioned earlier $\psi_{0}$ can be chosen real, which leads to the relations
$a_{1}=a_{-1}=a_{0}/2$ (see Appendix \ref{appendix:matrix-elements} for the details). This allows
to solve the recurrence relations and express all matrix elements in terms of a single real number
$a_0$. In addition, in Appendix \ref{appendix:matrix-elements} we consider a more complicated case
of matrix elements composed of functions that involve different representations.

Putting all contributions to the second-order response function together, we can
write it down in the following form:
\begin{align}
\nonumber
&
S^{(2)}=
\int d\zeta\frac{\partial\rho_0}{\partial E}
\sum\limits_{n=0}^{\infty}(-1)^n(a_n-a_{n+1})\left(n+\frac{1}{2}+s\right)\times
\\
\nonumber
&
\frac{\partial}{\partial t_1 }
\biggl(
\frac{\partial}{\partial t_2}
\Bigl(t_2\bigl(A_{n+1}(\zeta t_1)A_n^*(\zeta t_2)-A_n^*(\zeta t_1)A_{n+1}(\zeta t_2)\bigr)\Bigr)
\\
\label{S2-final}
&
-
\Bigl((n+1)A_{n+1}(\zeta t_1)A_n^*(\zeta t_2)+nA_n^*(\zeta t_1)A_{n+1}(\zeta t_2)\Bigr)
\biggr)
\,.
\end{align}
Here we used the short notation $A_n(t)$ for $A_n(t;s)$ and the fact that the product
$(2n+1-2s)A^*_{n+1}(t_2;s)A_n(t_1;s)$ is real. We used the symmetries $A_{-n}(t)=A_n(t)$ and
$a_{-n}=a_n$
discussed in Section \ref{subsection:dynamics-on-circle} and Appendix
\ref{appendix:matrix-elements} to restrict the consideration to non-negative $n$.

We note that the coefficients $A^*_n(-t_2;s)$ at negative time $-t_2<0$ enter the expansion for the
backward evolution of $\psi_0(\bm x;s)$, which appears to be crucial for the calculation below. The
time reversal symmetry $A_n(-t;s)=(-1)^{-n}A_n(t;s)$ was also used in Eq.~({\ref{S2-final}}).

The result for the second-order response function (\ref{S2-final}) is written in the form of an
ordinary series of terms whose expressions are either completely known analytically (see Eq.
(\ref{Fourier-coefficient2})) or recurrently expressed via a single number $a_0$ that characterizes
function $\psi_0(\bm x;s)$. The functions $\psi_{0}$ do not depend on the angle $\theta$, i.e.
$\psi_{0}({\bm x})=\phi({\bm r};\lambda)$, where, according to Eq.~(\ref{Casimir-Laplace})
$\phi({\bm r};\lambda)$ is an eigenmode of the Laplace operator $\nabla^{2}$ with the eigenvalue
$\lambda$ on our Riemann surface. Therefore, Eqs.~(\ref{measure-M2}) and (\ref{Casimir-Laplace})
allow to express the only undetermined parameters $s$ and $a_{0}$ in Eq.~(\ref{S2-final}) in terms
of the relevant eigenmode $\phi({\bm r};\lambda)$ of the Laplacian:
\begin{eqnarray}
\label{a0-s-Laplacian} a_{0}=\int_{M^{2}}d{\bm r}\left(\phi({\bm r};\lambda)\right)^{3}, \; s=
\frac{i\sqrt{-4\lambda-1}}{2}.
\end{eqnarray}

Finding the eigenmodes of the Laplacian analytically is a complex problem. Fortunately this is not
necessary for our purposes. It is known that for a given constant negative curvature $K<0$ there is
a family of non-equivalent Riemann surfaces of genus $g$ that can be parametrized by a space ${\cal
M}_{g}$ with the dimension ${\rm dim}{\cal M}_{g}=4g-2$ known as the moduli space. In particular,
$a_{0}$ and $s$ can be viewed as some nontrivial functions on the moduli spaces ${\cal M}_{g}$, and
computation of these functions is a complex problem. Yet, we can note that for any given values of
$a_{0}$ and $s$ we can find a Riemann surface that implements them according to
Eq.~(\ref{a0-s-Laplacian}). Therefore, we can treat $a_{0}$ and $s$ as free parameters that
characterize some Riemann surface of constant negative curvature.

In a more general case, when the dipole $f$ is represented by a finite superposition of $N_{f}$
Laplacian eigenmodes, the second-order response can be expressed in terms of a larger set of
parameters $s_{j}$ and $a_{0}^{s_{i}s_{j}s_{k}}$, with $i,j,k=1,\ldots,N_{f}$ as described in
Appendix \ref{appendix:matrix-elements}. We can apply the same argument to treat them as free
parameters for a Riemann surface of a high enough genus $g$.

\subsection{Second-order response: summation of the series}
\label{subsection:series-summation}

We have expressed the second-order response in terms of an ordinary converging series for the cases
of a single Laplacian mode $N_{f}=1$ [Eq.~(\ref{S2-final})] and two modes $N_{f}=2$
[Eq.~(\ref{S2-2modes})], respectively, in the expansion of the dipole (\ref{f-expansion}). A
general expression for a finite number $N_{f}$ of modes has a similar structure. Each term of the
series is known analytically in the form that involves special functions and solutions of
recurrence relations. Or goals are to derive the long-time asymptotics of the second-order response
$S^{(2)}$, and develop a computationally efficient procedure for $S^{(2)}(t_{1},t_{2})$ at finite
times.

We start with the asymptotic behavior of the second-order response function for large $t_1$ and
$t_2$. Naively, one would plug the long-time asymptotic of $A_n(t;s)$ from Eq.
(\ref{Fourier-coefficient-asympt}) into Eq. (\ref{S2-final}). However, this asymptotic of
$A_n(t;s)$ is valid only for fixed $n$, if $ne^{-t}\ll 1$. This can be confirmed by comparing the
consecutive terms in the expansion of the hypergeometric function in
Eq.~(\ref{Fourier-coefficient2}) in powers of $e^{-t}$.
Moreover, although $A_n(t;s)$ represent the Fourier expansion coefficients of a smooth function
(\ref{solution}), their asymptotic (\ref{Fourier-coefficient-asympt}) does seem to vanish as $n\to
\infty$. Counting the powers of $n$ in the second part of the summand in Eq. (\ref{S2-final})
estimated using these asymptotic expressions reveals that the resulting series fails to converge.

Indeed, the series (\ref{S2-final}) can be represented in the form
\begin{eqnarray}
\label{S-2-expand} S^{(2)}(t_{1},t_{2})=\sum_{n=0}^{\infty} (-1)^nF(n;t_{1},t_{2})\,,
\end{eqnarray}
where the dependence of $F(n;t_{1},t_{2})$ on $n$ is slow for $n\gg 1$. We may find the long-time
behavior of the terms as
\begin{eqnarray}
\label{term-double-expand}
F(n;t_{1},t_{2})=\sum_{kl=0}^{\infty}F_{kl}(n;t_{1},t_{2})e^{-2kt_{1}-2lt_{2}}\,.
\end{eqnarray}
The double expansion originates from the decomposition of $A_n(t;s)$ in powers of $e^{-2t}$. The
first term in the latter decomposition, specified by Eq.~(\ref{Fourier-coefficient-asympt}), turns
out to be independent of $n$ in the limit of large $n$, apart from irrelevant slow
quasi-oscillations $\propto n^s$ that cannot affect the series convergence. A naive long-time
asymptotic of the series is given by $\sum_n (-1)^nF_{00}(n;t_{1},t_{2})$, where the $n$-th term is
estimated as
\begin{align}
\label{F-00-explicit}
 & F_{00}(n;t_{1},t_{2})= n^{2}(a_n-a_{n+1})e^{-\frac{t_1+t_2}{2}}
 \nonumber \\
 &\times{\rm Re}\bigl(C_{n}(s)e^{st_1} \bigr) {\rm Re}\bigl(D_{n}(s)e^{st_2} \bigr).
\end{align}
where quasi-oscillatory dependence of $a_n$, $C_{n}(s)$ and $D_{n}(s)$ on $n\gg 1$ does not affect the
convergence. Taking into account the asymptotic form of $a_n$ [see Eq. (\ref{asymptotics})], we
conclude that the series is obviously divergent as $\sum_n (-1)^nF_{00}(n;t_{1},t_{2})\sim \sum_n
(-1)^n\sqrt{n}$. In what follows we show how to overcome the apparent divergence and calculate the
asymptotic of the second-order response function analytically.

The series in Eq.~(\ref{S-2-expand}) that represents the response function must converge for fixed
values of $t_{1}$ and $t_{2}$. Indeed, the terms $F(n;t_{1},t_{2})\propto
\exp[-2n(e^{-t_1}+e^{-t_2})]$ vanish exponentially, although only for extremely large $n\gg
e^{t_1},e^{t_2}\gg 1$. The decay rate is determined by the fact that $g(t,u)\equiv e^{-\hat L
t}\Psi_0(u)$ in Eq. (\ref{solution}) is smooth on scales that do not exceed the smallest fettuccine
size $e^{-t}$. The ultimate convergence allows to safely regroup the terms of the series:
\begin{eqnarray}
&&
\label{regrouped-series}
\sum\limits_{n=0}^{\infty}(-1)^nF(n)=
\frac{1}{2}F(0)+
\\
\nonumber
&&
\frac{1}{2}\sum\limits_{k=0}^{\infty}\bigl(F(2k)-2F(2k+1)+F(2k+2)\bigr)\,.
\end{eqnarray}
After the terms are regrouped the initial naive approach based on approximating $F(n;t_{1},t_{2})$
by its long-time asymptotic $F_{00}(n;t_{1},t_{2})$ results in a converging series, since the
linear combination in the summand represents the discrete counterpart of the second derivative
$d^2\, F_{00}(n)/dn^2$ which decays as $\propto n^{-3/2}$. According to Eq.~(\ref{F-00-explicit})
all terms in the resulting series have the same time dependence. Therefore, the long-time
asymptotic of the response function $S^{(2)}(t_{1},t_{2})$ is represented by a superposition of
four components $e^{-(t_1+t_2)/2\pm s(t_1\pm t_2)}$ with the coefficients in the form of convergent
series.

Coefficients  $F_{kl}(n;t_{1},t_{2})$ at higher orders in the expansion (\ref{term-double-expand})
grow faster with increasing $n$ as  $F_{kl}(n;t_{1},t_{2})\sim n^{2(n+k)}F_{00}(n;t_{1},t_{2})$.
The series re-grouping approach of Eq.~(\ref{regrouped-series}) can be generalized to eliminate the
apparent divergences for higher-order terms in Eq.~(\ref{term-double-expand}), and obtain the
higher-order terms in the asymptotic expansion of the second-order response in powers of $e^{-2t_{1,2}}$.
Instead of that we suggest an alternative procedure that allows (i) to demonstrate the existence of
a long-time asymptotic expansion of $S^{(2)}$ in powers of $e^{-2t_1}$ and $e^{-2t_2}$, (ii) derive
relatively simple expressions for the expansion coefficients in any order, and (iii) develop an
efficient numerical scheme for computation of the response at finite times.

We start with deriving the asymptotic expansion. To that end, provided $t_{1},t_{2}\gg 1$, we
introduce an intermediate $N\gg 1$, so that for $n>N$ the terms $F(n;t_{1},t_{2})$ are represented
by smooth functions of $n$. We further partition the sum $S$ of the series into the finite sum
$S_{N}=\sum_{n\le N}(-1)^{n}F(n)$ and the remainder $R_{N}=\sum_{n>N}(-1)^{n}F(n)$, followed by
evaluating the remainder.
Implementing the definition \cite{specfunctions}
of the Euler polynomials $E_{n}(y)$, we derive the following identity,
based on the Taylor expansion of $(1+e^{-x})^{-1}$ in $x=d/dz$:
\begin{align}
\nonumber
& F(z+1)-F(z)
= \sum\limits_{n=0}^\infty\frac{E_n(1)}{2n!}
\frac{d^n}{dz^n}(F(z+1)-F(z-1)),
\end{align}
which is
used to calculate the sum of consecutive terms pairwise, so that the remainder
$R_{N}=\sum_{n>N}(-1)^nF(n)$ of the almost alternating series becomes related to the first term
$F(N)$ that is not included in $R_{N}$, and its derivatives:
\begin{align}
\nonumber
&
R_{N}=
\frac{(-1)^{N+1}}{2}\sum\limits_{m=0}^\infty\frac{E_m(1)}{m!} F^{(m)}(N)=
\\
\nonumber
 &
 \frac{(-1)^{N+1}}{2}\sum\limits_{m=0}^{n}\frac{E_m(1)}{m!} F^{(m)}(N)
 + O\left(
 F^{(n+1)}(N)\right)=
\\
\label{remainder-expansion}
& (-1)^{N+1}\left(\frac{1}{2}F(N)+\frac{1}{4}F'(N)
+\dots \right) \,.
\end{align}
We further introduce the following ``improved'' partial sums:
\begin{align}
\label{PnN} & P^{(M)}_{N}=
 S_{N}+
 \frac{(-1)^{N+1}}{2}\sum\limits_{m=0}^{M}\frac{E_m(1)}{m!} F^{(m)}(N)\,,
\end{align}
so that $S^{(2)}=P^{(M)}_{N}+O(F^{(M+1)}(N))$. Due to the smoothness of $F(n)$ the deviation of
$S^{(2)}$ from its improved approximation $P^{(M)}_{N}$ may be estimated as $\sim F(N)/N^{(M+1)}$.
Since $P_{N}^{(M)}$ are determined by $F(z)$ for $z<N$, a choice of $N\ll e^{t_1},e^{t_2}$ allows
to use the expansion of Eq.~(\ref{term-double-expand}) which leads to an expansion of $P^{(M)}_{N}$
\begin{eqnarray}
\label{P-expand-time} P_{N}^{(M)}(t_{1},t_{2})=\sum_{m_{1}m_{2}}P_{N,m_{1}m_{2}}^{(M)}(t_{1},t_{2})
e^{-2(m_{1}t_{1}+m_{2}t_{2})}
\end{eqnarray}
in powers of $e^{-2t_1}$ and $e^{-2t_2}$ with
\begin{align}
\label{P-Nmm}
 & P^{(M)}_{N,m_{1}m_{2}}=
 \sum_{k=0}^{N}(-1)^{k}F_{m_{1}m_{2}}(k;t_{1},t_{2})+
 \\
 \nonumber
 &
 \frac{(-1)^{N+1}}{2}\sum\limits_{m=0}^{M}\frac{E_m(1)}{m!}
 \partial^m_N F_{m_{1}m_{2}}(N;t_{1},t_{2})\,.
\end{align}

The reason for introducing the improved partial sums $P^{(M)}_{N}$ is that the expressions for the
expansion coefficients $P^{(M)}_{N,m_{1}m_{2}}$ have finite limits at $N\to\infty$ provided
$M>m_1+m_2$. Stated differently, the second term in Eq.~(\ref{P-Nmm}) can be viewed as a set of
counter-terms that eliminate the divergence in the first term. In the limit $N\to\infty$
Eq.~(\ref{P-expand-time}) leads to the expansion
\begin{align}
\label{S-expand-time} S^{(2)}(t_{1},t_{2})=\sum_{m_{1}m_{2}} S^{(2)}_{m_{1}m_{2}}(t_{1},t_{2})
e^{-2(m_{1}t_{1}+m_{2}t_{2})}
\end{align}
with
\begin{align}
\label{Smm-limit}
 S^{(2)}_{m_{1}m_{2}}(t_{1},t_{2}) = \lim_{N\to\infty} P^{(m_1+m_2)}_{N,m_{1}m_{2}}(t_{1},t_{2})\,.
\end{align}

The expansion of Eq.~(\ref{S-expand-time}) is represented by an asymptotic rather than converging
series. It can be viewed as a double expansion of the second-order response function
$S^{(2)}(t_{1},t_{2})$ in Ruelle-Pollicott resonances which originates from the uncoupled
evolution of the chaotic system during time intervals $t_1$ and $t_2$. This will be demonstrated
explicitly \cite{CMResonances} by analyzing the noiseless limit of the corresponding Langevin
dynamics.

At this point we should note that our derivation of the asymptotic expansion has been somewhat
frivolous. First of all, since the terms $F(n;t_{1},t_{2})$ of our original series involve the
coefficients $a_{m}$, we do not actually know a function $F(z;t_1,t_2)$ that represents the series,
only its values for positive integer integer $z=n$ are available. In particular the derivatives
$\partial_{N}^{m}F_{m_{1}m_{2}}(N)$ in Eq.~(\ref{P-Nmm}) have been not defined yet. Second, in
obtaining the asymptotic expansion we were not controlling the neglected terms. This is especially
dangerous in our case since we derive the complete asymptotic, which involves computing
the terms that are exponentially small compared to more senior terms in the expansion. We need to
make sure that the terms that are neglected in computing a certain expansion coefficient are small
also compared to the higher terms that are kept in the asymptotic expansion.

A sketch of the appropriate derivation is presented in Appendix \ref{appendix:asymptotic}. In
particular, we demonstrate that the coefficients $a_n$ that enter the expressions for
$F(n;t_1,t_2)$ can be represented by some analytical function $a(z)$, and therefore the terms
$F(n;t_1,t_2)$ of the original series are represented by some analytical function $F(z;t_1,t_2)$.
We will further demonstrate how to compute derivatives $F^{(m)}(N)$ without explicit knowledge of
the function $F(z)$. We also show how a proper choice of $N$ and $n$ in $P^{(M)}_{N}$ allows
to control terms neglected in the asymptotic expansion.

In the remainder of this subsection we will implement the summation procedure in the form of an
efficient numerical scheme for computing the second-order response $S^{(2)}(t_1,t_2)$ at arbitrary
times. We reiterate that all terms in the series
for $S^{(2)}$ are known in their analytical form (via recurrence relations) apart from few
parameters determined by the geometry of the surface, as discussed at the end of the previous
subsection and in Appendix~\ref{appendix:matrix-elements}.

In the simplest case $N_{f}=1$ the single parameter is given by Eq.~(\ref{a0-s-Laplacian}). The
series for $S^{(2)}$ is absolutely convergent, which becomes remarkable only at $n\gtrsim
e^{t_{1}}$ or $n\gtrsim e^{t_{2}}$, and therefore a straightforward computation involves an
exponentially growing with time number of terms with relatively complex structure.
At given times $t_{1}$ and $t_{2}$ the terms $F(n;t_{1},t_{2})$ of the series constitute
a sequence of numbers.
Our numerical scheme
uses the procedure described above
and reduces the problem to the summation of a
relatively small and independent of time number of terms. This allows to compute
$S^{(2)}(t_1,t_2)$ with minimal numerical effort.

We can justify the numerical procedure separately for two overlapping intervals of
$t_{1}$ and $t_{2}$.
In the case $\min(t_{1},t_{2})\lesssim 1$ the series can be truncated already at $N\sim 10$.
At $n\gtrsim 10$ the decay of the terms $F(n;t_{1},t_{2})$ is exponential,
and the remainder is negligible.
Therefore, the numerical value of $S^{(2)}$ is obtained by the summation of several terms.

In the case $\min(t_{1},t_{2})\gtrsim 1$ we are not far from the asymptotic region $t_{1},t_{2}\gg
1$, and $S^{(2)}$ may be found by a simple numerical implementation of the remainder calculation
(\ref{remainder-expansion}). Since we use few terms in the expansion (\ref{remainder-expansion})
and approximate them by finite differences, the numerical precision is determined by how smoothly
the numbers $F(n;t_{1},t_{2})$ behave as $n$ increases. This is essentially determined by the value
of $n$. The smoothness
is only slightly influenced by the values of $t_{1},t_{2}\gtrsim 1$, which can be rationalized by
considering the principal asymptotic of $S^{(2)}$ at $t_{1},t_{2}\gg 1$. The series
(\ref{S-2-expand}) is not purely alternating because of the presence of quasi-oscillations $\propto
n^{\pm s}=e^{\pm i|s|\ln n}$ in $F(n;t_{1},t_{2})$. To ensure a sufficiently smooth behavior of
$F(n;t_{1},t_{2})$ at $n>N$ one must choose larger $N$ for larger values of $|s|$.

In practice, a reasonable relative error $\lesssim 10^{-3}$ is achieved if
merely $N\sim 10^2$ or less terms are retained
in the partial sum $S_{N}$ for $s\sim 5i$.
In the expansion (\ref{remainder-expansion}) the third term identically vanishes because $E_2(1)=0$.
We use two first terms in the expansion, and represent the first derivative
by the discrete difference with the third-order accuracy.
Therefore, the neglected contributions scale at best as
$O(F_{00}(N;t_{1},t_{2})e^{-4\min(t_{1},t_{2})})$. The efficiency of the procedure is remarkable.
We need to calculate only around $N\sim 10^2$ terms, which should be compared with a much bigger
number $e^{10}> 10^4$ at which the summand $F(n;t_{1},t_{2})$ starts to decay exponentially if
$\min(t_{1},t_{2})\sim 10$. At the first sight, summation of $10^2$ terms versus $10^4$ does not
seem to be a crucial improvement, given today's computational capacities. However, the bottleneck
here is computing the individual terms that are represented via the hypergeometric functions. The
computational effort grows dramatically for the terms $F(n;t_{1},t_{2})$ where the series starts to
converge in the absolute sense. Besides, straightforward summation of alternating series is not
free from accuracy issues. Overall, generating a $2D$ spectroscopic signal, with the developed
approach, requires several minutes of CPU time using such a simple and high-level tool as
Mathematica. This qualifies for an ``almost analytical'' calculation of the complete signal.

In conclusion, we emphasize that only a finite number of the series terms is used in the summation
procedure. Thus, the asymptotic expansion of the second-order response function turns out
to be determined by the expansions of $A_n(t;s)$ at $t\to\infty$. The latter may be viewed as
spectral decompositions of the evolution operator matrix elements. In fact, the strongly chaotic
system we consider is characterized by Ruelle-Pollicott resonances $\omega_{\nu}=\nu+1/2\pm s$.
Only even resonances with $\nu=2k$, where $k=0,1,\ldots$, contribute to the spectral decomposition.
This result for the second-order response is nontrivial because
the expansion can be made only after the summation of the series over angular harmonics.
This means that the expansion is only asymptotic.

The convergence issues can be completely avoided by introducing infinitesimal noise. Nonzero noise
provides the convergence of the series over angular harmonics for a given pair of resonances
\cite{CMResonances}. In the limit of the vanishing diffusion coefficient, the terms of the series
turn into their noiseless counterparts introduced in Eq. (\ref{term-double-expand}). The
application of the remainder summation procedure, described above, requires only smoothness of the
series terms. Therefore, the sum of the series is determined by a finite number of terms, and the
asymptotic expansion appears independent of the vanishing diffusion coefficient. This noise
regularization is meaningful, since the way the convergence is enforced is irrelevant for the
asymptotic expansion.

\section{Numerical results}
\label{section:numerical}

Experimental data on $2D$
time-domain spectroscopy that probes the response function $S^{(2)}(t_{1},t_{2})$ is usually
presented using the so-called $2D$ spectrum which is obtained by a numerical $2D$ Fourier transform of
the response function with respect to $t_{1}$ and $t_{2}$:
\begin{eqnarray}
\label{2dFourierEq}
S^{(2)}(\omega_1,\omega_2)=\iint\limits_0^\infty dt_1\,dt_2\, e^{i(\omega_1 t_1+\omega_2 t_2)}
S^{(2)}(t_1,t_2)\,.
\end{eqnarray}
To provide a spectroscopic view of chaotic vibrational dynamics, in this section we present
some numerical results on $2D$ spectra for a particular strongly chaotic system studied in this paper.

As we have shown above, the $2D$ response in our model can be expressed in terms
of the properties of the Laplacian eigenmodes that participate in the expansion
of the dipole function $f(\bm r)$.
Since the dipole is
generally a smooth function of the system coordinates, the number $N_f$ of relevant eigenmodes
is typically small.
To study the important features of the signals
we consider the second-order response function
in the simplest cases $N_f=1$ and $N_f=2$.
The latter represents the simplest situation that involves diagonal and cross peaks in the $2D$ spectrum.
Analytical results for this case which generalize Section \ref{section:calculation} are derived
in Appendix \ref{appendix:matrix-elements}.

In the case $N_f=1$ the second-order response function depends on a spectral parameter $s$
and an additional parameter $a_{0}$. Both parameters are expressed in terms of the only Laplacian eigenmode
that represents $f(\bm r)$ [see Eqs. (\ref{linear-response-result-1}), (\ref{a0-s-Laplacian}) and
(\ref{Casimir-Laplace})].

For $N_f=2$ we have two spectral parameters $s_1$ and $s_2$, and four additional parameters.
Similarly to the $N_f=1$ case, all parameters can be expressed in terms of the two relevant
Laplacian eigenmodes. As explained at the end of Section \ref{subsection:series-summation},
the parameters can be considered as independent attributes of a particular Riemann surface.
For the demonstration of qualitative features, we use a particular choice of the
parameters.

\begin{figure}[ht]
\centerline{
\includegraphics[width=2.8in]{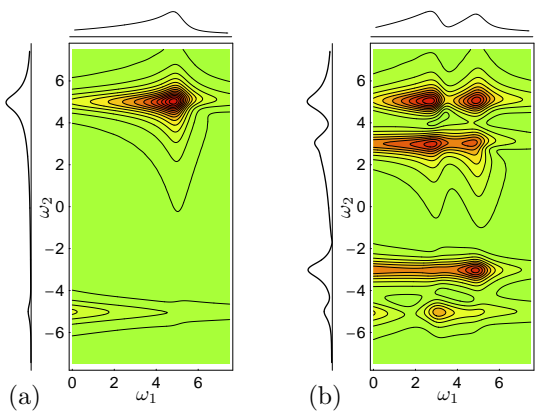}
}
\caption{
Absolute value of $2D$ Fourier transform of the second-order response function:
(a) single resonance $s=5i$, (b) linear combination of terms with two resonances $s_1=5i$ and $s_2=3i$.
Linear plots show cross-sections of the spectra
at $\omega_1=\omega_2=5$.
\label{2dFourier}
}
\end{figure}

\begin{figure}[ht]
\centerline{
\includegraphics[width=2.5in]{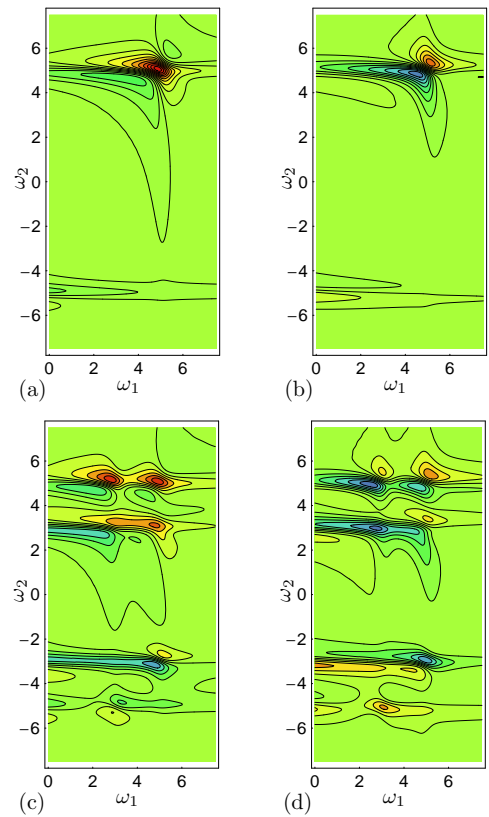}
}
\caption{
Real and imaginary parts of $2D$
spectra: (a) real part, (b) imaginary part with the single resonance $s=5i$; (c) real part, (d)
imaginary part of linear combination of terms with two different resonances $s_1=5i$ and $s_2=3i$.
\label{realimaginary}
}
\end{figure}

The absolute value of the $2D$ spectroscopic signal $|S^{(2)}(\omega_1,\omega_2)|$
is presented in Fig. \ref{2dFourier}.
Panel (a) corresponds to the $N_f=1$ case with the only spectral parameter $s=5i$.
Panel (b) shows the $N_f=2$ case with two spectral parameters $s_1=5i$ and
$s_2=3i$.
The real and imaginary components of the $2D$ spectra are presented in Fig. \ref{realimaginary}.
In the first case we see a diagonal peak with a pronounced stretched feature along $\omega_{1}$
direction.  In the second case we also see cross peaks accompanied by similar stretched features.

Diagonal and off-diagonal peaks are also observed in spectroscopic signals for harmonic and almost
harmonic vibrational dynamics. In this case the positions of the peaks are given by the frequencies
of the underlying periodic motions, whereas the width is determined by the system-bath interactions
and has about the same value in both frequency directions. In the chaotic case the peak positions
are determined by Ruelle-Pollicott resonances, rather than by the frequencies of some specific
periodic orbits. The stretching in spectroscopic signals originates from time-domain damped
``breathing'' oscillations with a variable period caused by strong nonlinearity of the underlying
vibrational dynamics and can be interpreted as a signature of chaos.

\section{Conclusion}
\label{section:discussion}

In the present paper we studied classical response in a strongly chaotic system. Using the
Liouville space representation for classical dynamics we found that the response functions exhibit
damped oscillations with time. The decay is attributed to the mixing property of the strongly
chaotic dynamics which leads to the efficient equilibration in phase space.

Previous studies of the integrable systems have revealed the appearance of unphysical divergences
in the classical nonlinear response functions at long times.
The divergence originates from the linear time growth of stability matrix elements in
integrable systems. Chaotic dynamics is characterized by the exponential growth of certain
stability matrix elements.

We have described a general qualitative picture of response in strongly mixing (hyperbolic)
systems. The exponential growth of the stability matrices that can potentially lead to exponential
divergence in response functions has been interpreted using the Liouville space representation of
classical mechanics: As a result of evolution a smooth initial distribution becomes extremely
sharp along the stable directions. The second interaction with the driving field involves
derivatives of the evolved distribution along the stable directions. This could result in
exponentially growing terms in nonlinear response functions. We have demonstrated that due to
smoothness of the initial distribution the dangerous terms completely cancel out, and the nonlinear
response functions exhibit exponential decay at long times. Stated differently, in mixing systems
stability matrices have exponentially growing and exponentially decaying components. Due to the
smooth character of the dipole function that describes the coupling to the driving field, the
growing components of the stability matrices are simply eliminated from the game, and the physical
behavior of the nonlinear response functions is determined by the exponentially decaying
components.

To confirm the established qualitative picture we performed detailed calculation of the linear and
second-order response for a chaotic model of a free particle moving along a compact surface with
constant negative curvature. The model possesses dynamical symmetry that allows for an exact
solution by applying the group representation theory. We found the long-time asymptotic behavior of
the linear and second-order response function that has a form of exponentially damped oscillations.
Complete asymptotic series obtained in this paper can be viewed as expansions in Ruelle-Pollicott
resonances. The expansion of the linear response in RP resonances is in agreement with the earlier
results for the two-time correlation functions \cite{RobertsMuz}, since the latter is directly
related to the linear response via the FD theorem.

The analogue of the RP expansion for the nonlinear response is of more nontrivial nature. The
eigenmodes that represent the RP resonances are generalized, rather than smooth functions. In the
linear case the initial smooth distribution should be decomposed in the RP modes. The signal is
computed by convoluting the RP modes with the smooth dipole function. Both operations are
well-defined for generalized functions. In the case of nonlinear response the second interaction
with the driving field involves acting on a generalized function with a differential operator
followed by projecting it onto a generalized function. The legitimacy of the latter operation is less
obvious, and is related to aforementioned cancellations of the dangerous terms. The RP
decomposition for the linear response is represented by a converging series, whereas the nonlinear
response is given by an asymptotic series. Computation of the expansion
coefficients in the nonlinear case requires a delicate summation of almost sign alternating series.
These are other signatures of the nontrivial character of the RP decomposition in the nonlinear
case.

In our forthcoming publication \cite{CMResonances} we consider Langevin dynamics associated with
the model considered here. Spectral decomposition in the corresponding eigenmodes of the
Fokker-Planck operator is a stable legitimate procedure. The spectral decomposition in the presence
of noise is a converging series. We show that in the limit of vanishing noise the converging series
becomes asymptotic and reproduces the expansion derived in this paper.
The asymptotic expansions of the linear and nonlinear response functions can be interpreted
as decompositions in RP resonances.

We suggest to apply our results for the interpretation of spectroscopic data. Classical chaos is
quite generic for large molecules. Even for smaller number of nuclear degrees of freedom, e.g. in
small systems of hydrogen bonds \cite{Hbonds}, the shape of the effective potential energy can make
the dynamics chaotic. We have argued that restricted motion in the regions with negative curvature
of the molecular potential can be qualitatively described as free motion along a compact Riemann
manifold with negative curvature. Moreover, the chaos may be caused even by the positive but
inhomogeneous curvature \cite{Pettini} of the configuration space.

We have considered in detail the
second-order response which is absent in the spectroscopy of the bulk materials or in isotropic
environments. Spectroscopy on the surface \cite{Ullman}, in the absence of the inversion symmetry,
can measure nonvanishing response functions of even order.

A dynamical system of the general type has the mixed phase space which includes both chaotic
regions and stability islands. Even in such systems, when the true long-time asymptotic decay of
correlations is rather power-like than exponential, RP resonances may by noticeable, leading to the
intermediate asymptotic, exponential decay persisting for a very long time \cite{FishmanRahav}.

When $2D$ spectroscopic data is interpreted in terms of the underlying dynamics the peaks are
usually attributed to some periodic motions in the system. We have shown that chaotic dynamics can
result in similar diagonal and off-diagonal peaks in spectroscopic signals that should be
attributed to RP resonances, and are nor related to any specific periodic orbits.  Note that the
frequencies and decay rates of the RP resonances can be retrieved from the dynamical
$\zeta$-function that can be represented as a product over all periodic orbits \cite{Ruelle86}.
Stated differently, the RP resonances are related to periodic orbits, yet in an extremely collective
way, and may not be attributed to any particular periodic motions. Therefore, they can be referred
to as collective chaotic resonances. Our numerical results show pronounced stretched features
associated with diagonal and off-diagonal peaks. These features result from breathing damped
oscillations that originate from contributions of multiple RP resonances with similar oscillation
frequencies (imaginary parts) and different damping rates (real parts). Breathing oscillations are
known to be typical for strongly nonlinear dynamical systems. On the other hand, $2D$ spectra
in harmonic or almost harmonic systems coupled to a multi-mode harmonic bath usually show similar
peak patterns in both frequency directions. We suggest that the stretched versus non-stretched peak
shape could have a capacity of distinguishing between the collective versus individual nature of
resonances in spectroscopic signals.


\appendix

\section{Dynamical Symmetry of Geodesic Flows in Riemann Surfaces with constant Negative Curvature
and Representation Theory}
\label{appendix:symmetry}

This Appendix contains some basic aspects on the geometry of Riemann surfaces with constant
negative curvature that are necessary for employing the $SO(2,1)$ dynamical symmetry. This allows
to decompose the free-particle dynamics using irreducible representations and map the original
dynamical problem  onto some effective $1$-dimensional dynamics in a circle.

The fundamental group $\Gamma_{g}=\pi_{1}(M_{g}^{2})$ of a Riemann surface $M_{g}^{2}$ of genus $g$
which describes noncontractible closed paths is generated by $2g$ elements $a_{j},b_{j}$, where
${j=1,\ldots,g}$, with the only relation $\prod_{j=1}^{g}a_{j}b_{j}a_{j}^{-1}b_{j}^{-1}=1$. In the
case $g>1$ the compact surface is covered $H\to M_{g}^{2}$ by the hyperbolic plane that can be
implemented as a pseudosphere determined by the equation $y_{1}^{2}+y_{2}^{2}-y_{0}^{2}=-1$ with
$y_{0}>0$ embedded in the $3D$ Minkowski space with a metric $d\ell^2=-dy_0^2+dy_1^2+dy_2^2$ (other
well-known equivalent implementations include the Poincare disc and the upper complex half-plane).
The fundamental group acts freely in $H$, so that $M_{g}^{2}\cong \Gamma_{g}\backslash H$. The
reduced phase space $M_{g}^{3}$, considered as a bundle $S^{1}\to M^{3}\to M^{2}$, whose fibers are
unit velocity or momentum vectors, being pulled back to $H$ forms a bundle $S^{1}\to G\to H$, where
$G\cong SO(2,1)$ can be represented as a pseudo-orthogonal group. This can be visualized as
follows: $G\cong SO(2,1)$ consists of points in $H$ (positions) that can be considered as $3D$
vectors with norm $-1$ in the Minkowski metric, together with  unit velocity vectors that can be
interpreted as norm $1$ vectors orthogonal to the position vectors (with respect to the Minkowski
metric). Extending these two pseudo-orthonormal vectors to a pseudo-orthonormal basis set, we
interpret $G$ as the space of pseudo-orthonormal basis sets, the latter can be thought of as the
pseudo-orthogonal group $SO(2,1)$. Factorizing $G\cong SO(2,1)$ with respect to the right action of
its maximal compact subgroup $K\cong SO(2)$ we arrive at $H\cong G/K$. The action of $\Gamma_{g}$
in $H$ can be naturally extended to a left action of $\Gamma_{g}$ in $G$, which determines the
embedding $\Gamma_{g}\subset G$. This leads to convenient for our purposes representations
$M_{g}^{2}\cong \Gamma_{g}\backslash G/K$ and $M_{g}^{3}\cong \Gamma_{g}\backslash G$. In
particular this interprets the action of $G$ in $M_{g}^{3}$ as originating from right action of $G$
in itself.

The presented picture has a very transparent interpretation. We start with a Riemann surface
$M_{g}^{2}$ of genus $g>1$ with constant negative scalar curvature $K=-1$. As stated in section
\ref{section:free-partcile-2D} we have three canonical vector fields in $M_{g}^{3}$, i.e.
$\sigma_{z}=\partial/\partial\theta$, $\sigma_{1}$ that determines the geodesic flow (classical
dynamics of a free particle), and $\sigma_{2}=[\sigma_{1},\sigma_{z}]$. In the case of constant
curvature they form the Lie algebra $so(2,1)$ with respect to the vector field commutator. Explicit
expressions for the vector fields are presented in Section \ref{section:free-partcile-2D} and in
Appendix \ref{appendix:poisson}. This defines an action of $so(2,1)$ in $M_{g}^{3}$ that can be
naturally extended to $G\to H$, considered as the pull-back of $M_{g}^{3}\to M_{g}^{2}$ to $H$. The
algebra action in $G$ can be integrated to a group action, which implies that locally $G$ has a
structure of the universal Lie group associated with the algebra $so(2,1)$. Careful studying of the
global properties of $G$ shows that in fact $G\cong SO(2,1)$. This immediately implies that $H\to
M_{g}^{2}$ is equivalent to $SO(2,1)/SO(2)$ as a Riemann space, i.e. $H$ is the hyperbolic space.
This has a very important implication that the local structure of any Riemann surface $M_{g}^{2}$
with constant scalar curvature $K=-1$ is locally equivalent to the hyperbolic space $H$, whereas
$M_{g}^{3}$ is locally equivalent to $G$. In particular, this implies that all local quantities
such as the Laplacian operator $\nabla^{2}$ in $M_{g}^{2}$, the Poisson bracket $\omega$ and the
geodesic flow in $M_{g}^{3}$ can be actually computed in $H$ and $G$, followed by expressing them
using the canonical vector fields $\sigma_{l}$, $l=1,2,z$. This is how, e.g.
Eq.~(\ref{Poisson-bracket-2}) can be derived in a straightforward way.

Finally we note that for a Riemann surface with $g>1$ and not necessarily constant curvature the
metric determines a complex-analytical structure, and the universal cover $H\to M_{g}^{2}$ is
equivalent to the hyperbolic plane and preserves the complex-analytical structure. This implies
$M_{g}^{2}\cong \Gamma_{g}\backslash H$. Since the group of conformal diffeomorphisms of $H$
coincides with its isometry group $SO(2,1)$, we have $\Gamma_{g}\subset G$. This defines a metric
in $M_{g}^{2}$ that has constant curvature $K=-1$ and is conformally-equivalent to the original
one. Since that, denoting by ${\cal M}_{g}$ the moduli space of complex analytical structures for
genus $g$ we can describe a Riemann surface with constant curvature by a pair $(\eta,K)$ with
$\eta\in{\cal M}_{g}$ and $K<0$ being the scalar curvature.

The representations of $M^{3}$ and $M^{2}$ in terms of the groups $K,\Gamma\subset G$
\begin{align}
&
\label{M3-M2-factorized}
M^{3}\cong \Gamma\backslash G\,, \;\;\; M^{2}\cong \Gamma\backslash G/K
\cong \Gamma\backslash H\cong M^{3}/K\,,
\\
&
\nonumber
H\cong G/K
\end{align}
are very convenient for describing invariant integration measures. Since $G$, $K$, and $\Gamma$ are
unimodular groups, according to the theorem on invariant measures in homogeneous spaces
\cite{Kirillov,Lang,Williams}, there is a unique invariant measure in $\Gamma\backslash G$ that is
locally equivalent (in the sense of covering) to the Haar invariant measure in $G$. The latter
generates an invariant measure in $G/K$, which is locally equivalent to the measure in $M^{2}$
generated by the metric. Combined with the main property of invariant measures in homogeneous
spaces, this implies for any integrable function $g({\bm x})$
\begin{eqnarray}
\label{measure-decompose} \int_{M^{3}}d{\bm x}g({\bm x})= \int_{M^{2}}d{\bm r}\int_{K_{{\bm
r}}}\frac{d\theta}{2\pi}g({\bm r},\theta)\,.
\end{eqnarray}
This measure provides an invariant scalar product in the space ${\cal H}$ of functions in $M^{3}$.
Besides, implementing functions $\tilde{g}({\bm r})$ in $M^{2}$ as functions $g({\bm x})$ in
$M^{3}$ that do not depend on $\theta$, i.e. $\sigma_{z}g=0$, we have
\begin{eqnarray}
\label{measure-M2} \int_{M^{2}}d{\bm r}\tilde{g}({\bm r})=\int_{M^{3}}d{\bm x}g({\bm x})\,,
\end{eqnarray}
and, therefore ${\cal H}$ can be decomposed into a direct sum of irreducible representations of $G$
\begin{eqnarray}
\label{decompose-functions}
{\cal H}={\cal H}^{(0)}\oplus\bigoplus_{s\in{\rm Spec}(M^{2})}{\cal H}_{s}
\oplus\bigoplus_{n\in\mathbb{Z}}^{n\ne 0}m_{n}{\cal H}^{(n)}\,.
\end{eqnarray}
In Eq.~(\ref{decompose-functions}) ${\cal H}^{(0)}$ denotes the one-dimensional unit representation
that represents constant functions, ${\cal H}_{s}$ denote the principal series representations
(with purely imaginary values of $s\in{\rm Spec_0}(M^{2})$ where ${\rm Im}s>0$) and complimentary
series representations (with real values of $0\le s<1/2$), whereas ${\cal H}^{(n)}$ are the
discrete series representations with the integer factors $m_{n}\ge 0$ describing how many times a
representation participates in the decomposition.

\section{Algebra $so(2,1)$ and Poisson bracket}
\label{appendix:poisson}

In this Appendix we derive the Poisson bracket in the form of Eq.~(\ref{Poisson-bracket-2}) with
the commutation relations between the operators $\sigma_l$ ($l=1,2,z$) given by
Eq.~(\ref{so(2,1)-commutation}).

We start with the canonical form of the Poisson bracket
in terms of coordinates $r^i$ and conjugated momenta $p_i=\partial L/\partial \dot r^i$:
\begin{eqnarray}
\label{canonical-Poisson}
\omega=\frac{\partial}{\partial p_i}\otimes\frac{\partial}{\partial r^i}
-\frac{\partial}{\partial r^i}\otimes\frac{\partial}{\partial p_i}\,,
\end{eqnarray}
where we imply summation over repeated indices.

Since free motion conserves the absolute value of the momentum $\zeta$,
it is convenient to make another local choice of variables in the tangent bundle $TM^3$.
We represent the particle momentum in terms of $\zeta$ and $\theta$,
the polar angle which determines the momentum direction.
The algebra element $\sigma_z$ which generates momentum rotations
in $TM^3$ reads $\sigma_z=\partial/\partial\theta$.

The third vector field in the reduced phase space is described by the differential operator
\begin{eqnarray}
\sigma_2 \equiv [\sigma_1,\sigma_z]=
E_{in}g^{ik}\left(\frac{\partial\zeta}{\partial p_n}\frac{\partial}{\partial r^k}
-\frac{\partial\zeta}{\partial r^k}\frac{\partial}{\partial p_n}\right)
\\
\nonumber
+\frac{\partial}{\partial q^k}\left(E_{in}g^{il}p_l\right)
\left(\frac{\partial\zeta}{\partial p_k}\frac{\partial}{\partial p_n}
-\frac{\partial\zeta}{\partial p_n}\frac{\partial}{\partial p_k}\right)\,.
\end{eqnarray}

Next, we show that $\sigma_1,\sigma_2,\sigma_z$ form the algebra $so(2,1)$.
The coefficients of the first order linear differential operator $[\sigma_2,\sigma_z]$
include the metric tensor and its first derivatives.
A straightforward yet tedious calculation performed in the basis set of four vectors related
to the canonical variables $p_i,q^i$ results in the relation $[\sigma_2,\sigma_z]=-\sigma_1$.
A similar calculation yields the commutator $[\sigma_{1},\sigma_{2}]=-K(\bm r)\sigma_{z}$ which
contains configuration space curvature $K(\bm r)$ and is indeed expressed only via the
remaining operator $\sigma_{z}$.
In the case of the constant negative curvature, rescaling  $\sigma_1$ and $\sigma_2$ leads
to the $so(2,1)$ commutation relations:
\begin{eqnarray}
\nonumber
[\sigma_{1},\sigma_{2}]=\sigma_{z}\,,
\quad
[\sigma_{1},\sigma_{z}]=\sigma_{2}\,, \;\;\;[\sigma_{2},\sigma_{z}]=-\sigma_{1}\,.
\end{eqnarray}

A number of transformations allow to express the Poisson bracket (\ref{canonical-Poisson})
in terms of the operators
$\partial_\zeta,\sigma_1,\sigma_2,\sigma_z$.
The simple form
\begin{align}
&
\omega=\frac{\partial}{\partial \zeta}\otimes\sigma_1
- \sigma_1\otimes \frac{\partial}{\partial \zeta}
+\frac{1}{\zeta} \left( \sigma_2\otimes\sigma_z - \sigma_z\otimes\sigma_2 \right)
\end{align}
reflects the presence of the dynamical symmetry.

The first two terms of the Poisson bracket determine the phase space velocity.
The constant coefficient in front of the second term can be alternatively deduced
from the condition $S^{(2)}(t_1,0)=0$.

\section{Unitary Irreducible Representations of $SO(2,1)$}
\label{appendix:irreducible-SO(2,1)}

In this Appendix we describe a convenient implementation of unitary irreducible representation of
$SO(2,1)$ in terms of functions in a circle $S^{1}$. This provides an aforementioned mapping of
the free-particle dynamics onto a $1$-dimensional problem. Since the group $G\cong SO(2,1)$ has a
double covering $SL_{2}(R)\to SO(2,1)$, irreducible representations of $G$ are provided by even
irreducible representations of $SL_{2}(R)$, the latter being well-known (see e.g., \cite{Lang}).
This situation is conceptually close the the case of irreducible representations of $SO(3)$ given
by even, i.e. integer-spin, representations of a double cover $SU(2)\to SO(3)$ of
$SO(3)$. We follow the approach of \cite{Lang} for $SL_{2}(R)$ and translate it
to the language convenient for $G\cong SO(2,1)$.

The space ${\cal H}_{s}$ of an irreducible representation of $G$ that belongs to the principal or
complimentary series has a convenient basis set $\Psi_{k}$ of angular harmonics, so that
$\sigma_{z}\Psi_{k}=ik\Psi_{k}$. The anti-Hermitian conjugated raising and lowering operators
$\sigma_{\pm}=\sigma_{1}\pm i\sigma_{2}$,  Using the
allow for the the following relations between the basis functions:
\begin{align}
\label{sigma-act-irred}
&&
\sigma_{\pm}\Psi_{k}=\left(\pm k+\frac{1}{2}-s\right)\Psi_{k\pm 1}\,, \;\;\;
\sigma_{z}\Psi_{k}=ik\Psi_{k}\,,
\\
\nonumber
&&
\left(\Psi_{k},\Psi_{k'}\right)=0, \; {\rm for} \; k\ne k'\,.
\end{align}
In the case of imaginary $s$ (principal series) $\Psi_{m}(u)$ are
normalized, and correspond to normalized functions $\psi_{k}({\bm x};s)$, i.e.
\begin{eqnarray}
\label{orth-cond-M3}
\int_{M^{3}}d{\bm x}\left(\psi_{k'}({\bm x};s')\right)^{*}\psi_{k}({\bm x};s)=
\delta_{k'k}\delta_{s's}\,,
\end{eqnarray}
where $d{\bm x}$ is the invariant measure in $M^{3}$ defined up to a constant
\cite{Kirillov,Lang,Williams}.
We can alternatively view Eq.~(\ref{sigma-act-irred}) as a definition of a $so(2,1)$ representation
parametrized by $s$. In Appendix \ref{appendix:poisson} we directly verify
that the operators $\sigma_{\pm},\sigma_{z}$
satisfy the $so(2,1)$ commutation relations given by Eq.~(\ref{so(2,1)-commutation}). The
definition of Eq.~(\ref{so(2,1)-commutation}) becomes clear when we implement the representation
space ${\cal H}_{s}$ as a Hilbert vector space of functions ${\Psi}(u)$ in a
circle:
\begin{align}
\label{so(2,1)-circle}
&
\sigma_{z}=\frac{d}{du}, \quad \sigma_{1}=\sin u\frac{d}{du}+\frac{1-2s}{2}\cos u\,,
\\
&
\sigma_{2}=-\cos u\frac{d}{du}+\frac{1-2s}{2}\sin u\,,
\nonumber
\\
&
\sigma_{\pm}=\exp(\pm iu)\left(\mp i\frac{d}{du}+\frac{1-2s}{2}\right)\,, \ \
\Psi_{k}(u)=\exp(iku)\,.
\nonumber
\end{align}
The validity of Eqs. (\ref{so(2,1)-commutation}) and (\ref{sigma-act-irred}) for the generators
$\sigma$ and basis vectors $\Psi_{k}$ defined by Eqs.~(\ref{so(2,1)-circle}) can be verified
directly. The reason for such a simple representation of the generators in the form of first-order
differential operators is that each representations ${\cal H}_{s}$ is induced \cite{Kirillov} from
a one-dimensional representation, parametrized by $s$, of a two-dimensional subgroup $AN\subset G$
\cite{Lang}. It is implemented in the space of functions in the maximal compact subgroup $K\subset
G$ where $K\cong SO(2)\cong S^{1}$. In the case of imaginary $s$ (principal series)
the inducing representation is unitary. Therefore the induced
representation has a natural scalar product \cite{Kirillov}
\begin{eqnarray}
\label{scal-prod-circle}
\left(\Psi,\Psi'\right)=\int_{0}^{2\pi}\frac{du}{2\pi}\left(\Psi'(u)\right)^{*}\Psi(u)\,.
\end{eqnarray}
It can be easily verified that the generators $\sigma_{l}$ for $l=1,2,z$ in the implementation of
Eq.~(\ref{so(2,1)-circle}) are anti-Hermitian operators with respect to the scalar product defined
by Eq.~(\ref{scal-prod-circle}).

It also follows from Eq.~(\ref{scal-prod-circle}) that the set of
vectors $\Psi_{k}$ constitute an orthonormal basis with the positive scalar product
$\left(\sigma_{+}\Psi_{0},\sigma_{+}\Psi_{0}\right)=(1-4s^{2})/4$ for imaginary $s$.
Besides imaginary $s$, the scalar product is also positively defined
for real $s$ with $-1<2s<1$ (complimentary series).
In this case the representations ${\cal H}_{s}$ are not unitary with respect
to the scalar product of Eq.~(\ref{scal-prod-circle}),
however,
another scalar product still diagonal in the
basis set of $\Psi_{k}$ can be defined
(this procedure is also known as representation unitarization \cite{Lang}).
We do not use the the complimentary
series representations in the calculation of the response because they lead to the exponential decay
without the oscillatory behavior.

Irreducible representations ${\cal H}^{(n)}$ of the discrete series can be considered as
sub-representations ${\cal H}^{(n)}\subset {\cal H}_{n+1}$, and ${\cal H}^{(n)}\subset {\cal
H}_{n-1}$ for $n>0$ and $n<0$, respectively, generated by the vectors $\Psi_{k}$ with $k\ge n$ and
$k\le n$, respectively, which creates a certain inconvenience. This does not constitute a major
problem, since the discrete representations ${\cal H}^{(n)}$ can be alternatively holomorphically
induced from unitary representations of the maximal compact subgroup $K\cong SO(2)$. We are not
discussing this construction here, since discrete representations do not contain the zero-momentum
state $\Psi_{0}$, and, therefore, they do not contribute to the linear and second-order response.

It is also important to note that irreducible representations of the principal and complimentary
series characterized by opposite values of $s$ are unitary equivalent, i.e.
${\cal H}_{-s}\cong{\cal H}_{s}$,
which follows from the fact that the representation unitarization is
determined by the value of $s^{2}$ \cite{Lang}. For the principle series representations
this property is
clearly seen from Eq.~(\ref{sigma-act-irred}). The latter implies that in the case of imaginary $s$ the
vectors $\left(\sigma_{\pm}\right)^{k}\Psi_{0}$ for opposite values of $s$ are different by just
phase factors and can be connected by a unitary transformation that is
diagonal in the $\Psi_{k}$ basis set. The purpose of the unitary transformation is
to compensate the aforementioned phase factors.
In particular, this justifies the agreement that the spectrum ${\rm Spec}(M^{2})$ contains only
imaginary $s$ with ${\rm Im}s>0$ and real $s$ with $0\le s<1/2$.

We conclude this appendix by relating the spectrum ${\rm Spec}(M^{2})$ to the spectrum of the
Laplacian in $M^{2}$. To that end we introduce the Casimir operator $\hat{C}$ that commutes with
all $so(2,1)$ generators, and, therefore, is a constant in any irreducible representation due to
the Shur lemma
\begin{align}
\label{Casimir}
&
\hat{C}=-\frac{1}{2}(\sigma_{+}\sigma_{-}+\sigma_{-}\sigma_{+})+\left(\sigma_{z}\right)^{2}\,,
\\
\nonumber
&
[\hat{C},\sigma_{l}]=0 \; {\rm for} \; l=1,2,z\,,
\quad
\hat{C}\Psi=\frac{1-4s^{2}}{4}\Psi \; {\rm for} \;
\Psi\in{\cal H}_{s}\,,
\end{align}
where we used ladder operators $\sigma_\pm$.
We can further interpret functions in $M^{2}$ as functions $f(\bm x)$ in $M^{3}$ independent of the
momentum direction $\theta$, i.e. $\sigma_{z}f=0$. It is one of the signatures of the dynamical
symmetry that,
if acting on functions in $M^{2}$, the Laplacian
operator (which is proportional to the Hamiltonian of a quantum free particle) is expressed in terms of
the Casimir operator as
\begin{eqnarray}
\label{Casimir-Laplace}
&&
-\nabla^{2}=-\frac{1}{2}(\sigma_{+}\sigma_{-}+\sigma_{-}\sigma_{+})=\hat{C}-\left(\sigma_{z}\right)^{2}\,,
\\
\nonumber
&&
\nabla^{2}\psi_{0}({\bm x};s)=-\frac{1-4s^{2}}{4}\psi_{0}({\bm x};s)\,.
\end{eqnarray}
It follows from Eq.~(\ref{Casimir-Laplace}) that the spectrum ${\rm Spec}_{0}(M^{2})$ of a Riemann
surface is totally determined by the spectrum of its Laplacian $\nabla^{2}$. The spectrum of the
Laplacian on a compact surface is discrete, and the eigenvalues are negative.
The functions
$\psi_{k}({\bm x};s)$ with $s\in{\rm Spec}_{0}(M^{2})$ that constitute a basis set in the space of
relevant distributions according to Eqs. (\ref{decompose-distribution}) and (\ref{decomp-ang-mom})
can be expressed in terms of the Laplacian eigenfunctions $\psi_{0}({\bm x};s)$  using
ladder operators according to relations (\ref{sigma-act-irred}).

For the sake of completeness we note that the functions $\psi_{k}({\bm x};s)$ for $s\in{\rm
Spec}_{0}(M^{2})$ and fixed value of $k$, known as modular forms of degree $k$ (see, e.g.
\cite{Lang}) are eigenmodes of the Laplacian operator $\nabla^{2}$ still defined by
Eq.~(\ref{Casimir-Laplace}).
The Laplacian operator can be viewed as the quantum Hamiltonian of a free particle
moving in homogeneous magnetic field, whose intensity is proportional to $k$.

Given the function $\Psi(u)$ in the circle, one can find its counterpart
$\psi(\bm x;s)$ in representation space ${\cal H}_s$:
\begin{align}
\psi(\bm x;s)=\sum\limits_{k=-\infty}^{\infty}
\int\limits_0^{2\pi} du\, e^{-iku} \Psi(u) \psi_k(\bm x;s)\,,
\end{align}
where $\psi_k(\bm x;s)$ are basis functions in ${\cal H}_s$.

\section{Matrix elements in second-order response function}
\label{appendix:matrix-elements}

This Appendix includes some details necessary for the calculation of the
second-order response. The latter contains integrals of triple products like
$\psi^*_n(\bm x;s_1)(\sigma_{1,2}\psi_0(\bm x;s_2))\psi_m(\bm x;s_2)$. The Appendix describes
this geometrical part in the
calculation of the second-order response function, whereas the dynamical part is mapped
onto the problem in the circle and solved Section \ref{section:free-partcile-2D}.
The
anti-Hermitian operators $\sigma_l$ ($l=1,2,z$) play a fundamental role in the description
of the dynamics.
They are associated with vector fields that commute according to Eqs. (\ref{so(2,1)-commutation}).

In Section \ref{section:matrix-elements} we have derived the recurrence relations for matrix elements
in the simplest case $s_1=s_2=s_3=s$.
Since $\psi_0$ is a real-valued function (see Section \ref{section:classical-liouville}),
we notice that
$a_1=a_{-1}=a_0/2$. The recurrence relation (\ref{recurrent}) is symmetric with respect to the sign reversal
of $k$ and we conclude that $a_{-k}=a_k$.
Therefore, all terms in sets $a_k$ and $b_k$ are determined by a single real number $a_0$.
Another way to come to the conclusion is to notice that
\begin{eqnarray}
\psi_{n}^*(\bm x,s)=
\frac{\Gamma\left(n+\frac{1}{2}-s\right)\Gamma\left(\frac{1}{2}+s\right)}
{\Gamma\left(n+\frac{1}{2}+s\right)\Gamma\left(\frac{1}{2}-s\right)}\psi_{-n}(\bm x,s)\,.
\end{eqnarray}

Next we calculate the coefficients involved in the second-order response in a general situation
when the coupling $f$
includes contributions from several irreducible representations characterized by imaginary numbers
$s_1$, $s_2$, \dots . In this case the coefficients are labeled by additional indices:
\begin{eqnarray}
\label{abqrs}
a^{qrs}_k=\int d\bm x\,\psi_k^*(\bm x;q)\psi_0(\bm x;r)\psi_k(\bm x;s)\,,
\\
b^{qrs}_k=2\int d\bm x\,\psi_k^*(\bm x;q)\sigma_+\psi_0(\bm x;r)\psi_{k-1}(\bm x;s)\,.
\end{eqnarray}
Choosing the zero momentum eigenfunctions $\psi_0^*(\bm x;q)$ to be real,
$a^{qrs}_0$ remains invariant under permutations in $\{q,r,s\}$.
Employing Eqs.~(\ref{sigma-act-irred})
the recurrence relations (\ref{recurrent}) can be generalized:
\begin{align}
\label{abqrs-recurrent}
&
(2k+1+2q)(2k+1-2s)a^{qrs}_{k+1}=
\\
\nonumber
&
-(2k-1-2q)(2k-1+2s)a^{qrs}_{k-1}
\\
\nonumber
&
+(8k^2+1-4q^2+4r^2-4s^2)a^{qrs}_{k}\,,
\\
&
b^{qrs}_{k+1}=(2k+1-2q)a^{qrs}_k-(2k+1-2s)a^{qrs}_{k+1}\,.
\end{align}
This is done similarly to the particular case $q=r=s$ described above.
The asymptotic form that generalizes Eq. (\ref{asymptotics}) consists
of two contributions:
\begin{eqnarray}
a^{qrs}_{k}\propto k^{-\frac{1}{2}-q+s\pm r}
\,.
\end{eqnarray}

The numerical results of Section \ref{section:numerical} are presented
for the case of two terms in the expansion (\ref{f-expansion}) of the dipole
moment $f$. This is the simplest case that involves cross-peaks in $2D$ spectroscopic signals.
The second-order response has a form similar to Eq. (\ref{S2-final}):
\begin{align}
\label{S2-2modes}
& S^{(2)} =\int d\zeta\frac{\partial\rho_0}{\partial E} \frac{\partial}{\partial
t_1}\sum\limits_{p,q,r=s_{1},s_{2}} B_p B_q B_r \sum\limits_{n=0}^{\infty} (-1)^n \times
\\
&
\nonumber
\Biggl(
\biggl(\Bigl(n+\frac{1}{2}+r\Bigr)a^{pqr}_n-\Bigl(n+\frac{1}{2}+p\Bigr)a^{pqr}_{n+1}\biggr)
\times
\\
&
\nonumber
\biggl(t_2\frac{\partial}{\partial t_2}-n\biggr)\biggl( A_{n}^*(t_2;p)A_{n+1}(t_1;r)\biggr)
\\
&
\nonumber
-\biggl(\Bigl(n+\frac{1}{2}-p\Bigr)a^{pqr}_n-\Bigl(n+\frac{1}{2}-r\Bigr)a^{pqr}_{n+1}\biggr)
\times
\\
&
\nonumber
\biggl(t_2\frac{\partial}{\partial t_2}+n+1\biggr)\biggl( A_{n+1}^*(t_2;p)A_{n}(t_1;r)\biggr)
\Biggr)
\,.
\end{align}

Since $q,r,s$ adopt only two values $s_1$ and $s_2$, the
coefficients turns out $a_k$ to be symmetric, $a^{qrs}_k=a^{qrs}_{-k}$,
and Eqs. (\ref{abqrs-recurrent}) at $k=0$ allow to express $a^{qrs}_1$ via $a^{qrs}_0$.
Since the zero harmonics $\psi_0(\bm x;s)$ are real, the coefficients $a^{qrs}_0$ are also real
and invariant under any permutation of $q$, $r$ and $s$.
Therefore,
all coefficients $a^{qrs}_k$ are divided into four groups and expressed via only four real numbers
$a^{s_1s_1s_1}_0$, $a^{s_2s_2s_2}_0$, $a^{s_1s_1s_2}_0$ and $a^{s_1s_2s_2}_0$.

The properties of the sequences $\{a^{s_1s_1s_1}_k\}$ and $\{a^{s_2s_2s_2}_k\}$ have been described above.
The conditions necessary to express all numbers in these sets via the terms at $k=0$
are given by $a^{sss}_1=a^{sss}_0/2$.

The third group of matrix elements includes $a^{s_1s_1s_2}_k$, $a^{s_1s_2s_1}_k$ and $a^{s_2s_1s_1}_k$.
We express all these matrix elements via $a^{s_1s_1s_2}_0$ using the following relations:
\begin{align}
&
a^{s_1s_1s_2}_k=\left(a^{s_2s_1s_1}_k\right)^*\,,
\quad
a^{s_1s_2s_1}_1=\frac{1+4s_2^2-4s_1^2}{2(1-4s_1^2)}a^{s_1s_1s_2}_0\,,
\\
&
a^{s_2s_1s_1}_1=\frac{1-2s_2}{2(1-2s_1)}a^{s_1s_1s_2}_0
\,.
\end{align}

The fourth group of matrix elements  includes $a^{s_2s_2s_1}_k$, $a^{s_2s_1s_2}_k$ and $a^{s_1s_2s_2}_k$
which obey the relations similar to those for the third group but with $s_1$ and $s_2$ interchanged.
Thus they expressed in terms of a single real number $a^{s_2s_2s_1}_0$.

We assume that four numbers
$a^{s_1s_1s_1}_0$, $a^{s_2s_2s_2}_0$, $a^{s_1s_1s_2}_0$ and $a^{s_1s_2s_2}_0$
to be independent.
This is rationalized in a detailed discussion at the end of subsection \ref{section:matrix-elements}.

\section{Asymptotic expansion}
\label{appendix:asymptotic}

In this appendix we present some details involved in a derivation of the asymptotic expansion given
by Eq.~(\ref{S-expand-time}).

First of all the derivation is based on the assumption that the the terms $F(n)$ in our series can
be represented by some smooth function $F(z)$. This is true for the coefficients $A_{n}$ that enter
Eq.~(\ref{S2-final}) due to their explicit form given by Eq.~(\ref{Fourier-coefficient3}). However,
the coefficients $a_{n}$ that are also a part of Eq.~(\ref{S2-final}) are given as a discrete set
according to Eq.~(\ref{recurrent}). To show that $a_{k}$ can be represented by a smooth function
$a(z)$ we note that since $a_{k}$ is a solution of a recurrence relation,
its $k\to\infty$ behavior can be
readily analyzed, which results in an expansion
\begin{eqnarray}
\label{asymptotics}
 a_{k}=k^{-\frac{1}{2} + s}\sum_{m=0}^{\infty}\frac{\tilde{a}_{m,s}}{k^{m}}+
 k^{-\frac{1}{2} - s}\sum_{m=0}^{\infty}\frac{\tilde{a}_{m,-s}}{k^{m}}\,,
\end{eqnarray}
which yields an analytical function $a(z)$ in a form of an analytical expansion
\begin{eqnarray}
\label{a(z)-expand-1/z}
 a(z)=z^{-\frac{1}{2} + s}\sum_{m=0}^{\infty}\frac{\tilde{a}_{m,s}}{z^{m}} +
 z^{-\frac{1}{2} - s}\sum_{m=0}^{\infty}\frac{\tilde{a}_{m,-s}}{z^{m}}
\end{eqnarray}
in powers of $1/z$ at $z=\infty$.

So far we have demonstrated that the terms $F(k)$ in our series are represented by an analytical
function $F(z)$, and therefore an approach for calculating the asymptotic expansion introduced in
Section \ref{subsection:series-summation} is in principle legitimate. However, this is not enough
yet, since the asymptotic expansion coefficients $S_{m_{1}m_{2}}^{(2)}$ involve the derivatives
$\partial_{z}^{m}F_{m_{1}m_{2}}(z)$ at integer $z=k$ values of the argument, as prescribed by
Eq.~(\ref{P-Nmm}). Apparently, it requires explicit knowledge on the analytical function $F(z)$.
However, what we really need as a good enough approximation for the derivatives in
Eq.~(\ref{P-Nmm}) so that the neglected terms vanish in the limit $N\to\infty$. This can be
achieved by implementing a very simple scheme. For an analytical function $F(z)$ denote by
$F_{m}^{(k)}(N)$ an $m$-th order approximation for its derivative $\partial_{z}^{k}F(z)|_{z=N}$ at
$z=N$. Then the derivatives $\partial_{N}^{k}F_{m_{1}m_{2}}(N)$ in Eq.~(\ref{P-Nmm}) can be safely
replaced by their approximate values $F_{m,m_{1}m_{2}}^{(k)}(N)$, provided $m$ is large enough. The
approximate derivatives can be linearly expressed in terms of the series terms $F(n)$ by solving an
obvious system of linear equations
\begin{eqnarray}
\label{find-approx-deriv} \sum_{l=0}^{m}\frac{n^{l}}{l!}F_{m}^{(l)}(N)=F(N+n), \;\;\; n=0,\ldots,m.
\end{eqnarray}

In the remainder of this appendix we show how to control the neglected terms in a derivation of the
asymptotic series. We start with noting that the asymptotic expansion of Eq.~(\ref{S-expand-time})
should be understood in a way that for any pair $(M_{1},M_{2})$ of positive integers we have
\begin{eqnarray}
\label{S-expand-time-rig} S^{(2)}(t_{1},t_{2})&=&\sum_{m_{1}=0}^{M_{1}}\sum_{m_{2}=0}^{M_{2}}
S^{(2)}_{m_{1}m_{2}}(t_{1},t_{2}) e^{-2(m_{1}t_{1}+m_{2}t_{2})}\nonumber \\
&+&o\left(e^{-2(M_{1}t_{1}+M_{2}t_{2})}\right)\,.
\end{eqnarray}
Full control over the neglected terms can be achieved by a proper choice of the intermediate
$N(t_{1},t_{2})$ as a function of $t_{1},t_{2}\to\infty$ and $n$ in Eq.~(\ref{P-expand-time}). Note
that both $N(t_{1},t_{2})$ and $n$ also depend on $M_{1}$ and $M_{2}$. We choose
\begin{eqnarray}
\label{define-N(t)} N(t_{1},t_{2})=\exp\left(\gamma\min(t_{1},t_{2})\right)\,,
\end{eqnarray}
so that for small enough $\gamma$ we have $1\ll N \ll e^{t_{1}},e^{t_{2}}$ and we can expand
$F(n;t_{1},t_{2})$ for $n\le N$ in powers of $e^{-2t_{1}}$ and $e^{-2t_{2}}$ according to
Eq.~(\ref{term-double-expand}).
If we keep the terms up to the degree $(M_{1},M_{2})$,
the neglected terms can be estimated as $\sim
N^{2(M_{1}+M_{2})+r}\exp\left(-2\min(t_{1},t_{2})\right)e^{-2(M_{1}t_{1}+M_{2}t_{2})}$, where
$r=3$ corresponds to a very loose estimate. By choosing
$\gamma$ to be small enough, specifically $\left(2(M_{1}+M_{2})+r\right)\gamma<2$, the neglected
terms are $o\left(e^{-2(M_{1}t_{1}+M_{2}t_{2})}\right)$ and can be safely omitted. Finally, no
matter how small $\gamma$ is we can always find $M$ to be large enough (and independent of $t_{1}$
and $t_{2}$), so that $N^{-M}(t_{1},t_{2})=o\left(e^{-2(M_{1}t_{1}+M_{2}t_{2})}\right)$, and the
terms omitted in Eq.~(\ref{P-Nmm}) can be also safely neglected.

\end{document}